\documentclass[prd,paper,onecolumn,amsmath,amsfonts,amssymb,nofootinbib]{revtex4}
\usepackage {amsmath}
\usepackage[dvips]{graphicx}
\usepackage{array}
\usepackage{feynmp}
\usepackage{array}
\usepackage{amssymb}
\newcommand{\be}{\begin{equation}}
\newcommand{\ee}{\end{equation}}
\newcommand{\bear}{\begin{eqnarray}}
\newcommand{\eear}{\end{eqnarray}}
\newcommand{\tanb}{\tan \beta}

\newcommand{\bra}[1]{\left\langle #1\right|} 
\newcommand{\ket}[1]{\left| #1\right\rangle} 
\newcommand{\vev}[1]{\left\langle #1\right\rangle}

\newcommand{\MeV}{\; \mathrm{MeV}} 
\newcommand{\GeV}{\; \mathrm{GeV}} 
\newcommand{\TeV}{\; \mathrm{TeV}} 
\newcommand{\lapproxeq}{\lower .7ex\hbox{$\;\stackrel{\textstyle  
<}{\sim}\;$}} 
\newcommand{\gapproxeq}{\lower .7ex\hbox{$\;\stackrel{\textstyle  
>}{\sim}\;$}} 
\newcommand{\stackdown}[2]{\lower 1.4ex\hbox{$\;\stackrel{\textstyle{#1}}  
{\scriptstyle{#2}}\;$}}
\newcommand{\beq}{\begin{equation}} 
\newcommand{\eeq}{\end{equation}} 
\newcommand{\ba}{\begin{eqnarray}}
\newcommand{\ea}{\end{eqnarray}}

\newcommand{\bea}{\begin{eqnarray}}
\newcommand{\eea}{\end{eqnarray}}

\makeatletter 
\def\slash{\@ifnextchar[{\fmsl@sh}{\fmsl@sh[0mu]}} 
\def\fmsl@sh[#1]#2{% 
  \mathchoice 
    {\@fmsl@sh\displaystyle{#1}{#2}}% 
    {\@fmsl@sh\textstyle{#1}{#2}}% 
    {\@fmsl@sh\scriptstyle{#1}{#2}}% 
    {\@fmsl@sh\scriptscriptstyle{#1}{#2}}} 
\def\@fmsl@sh#1#2#3{\m@th\ooalign{$\hfil#1\mkern#2/\hfil$\crcr$#1#3$}} 
\makeatother 
\begin{document}
\begin{flushright} 
\parbox{4.6cm}{UA-NPPS/BSM-1-12 }
\end{flushright}
\title{Dilaton dominance relaxes LHC and cosmological constraints in supersymmetric models }
\author{A. B.~\ Lahanas}
\email{alahanas@phys.uoa.gr}
\affiliation{University of Athens, Physics Department,  
Nuclear and Particle Physics Section,  
GR--15771  Athens, Greece}
\author{Vassilis~C.~Spanos}
\email{spanos@inp.demokritos.gr}
\affiliation{Institute of Nuclear Physics, NCSR ``Demokritos'', GR-15310 Athens, Greece}

\vspace*{2cm}
\begin{abstract}
It has been pointed out recently that the presence of dilaton field in the early Universe can dilute the 
neutralino dark matter (DM) abundance, if Universe is not radiation dominated at DM decoupling, due to its dissipative-like coupling to DM. In this scenario two basic mechanisms 
compete, the modified Hubble expansion rate tending to increase the relic density and a dissipative force that tends to decrease it. The net effect can lead to an overall dramatic decrease of the predicted relic abundance, sometimes by amounts of the order of $\, {\cal {O }} (10^2)  $  or so. This feature is rather generic, independent of any particular assumption on the underlying string dynamics, provided dilaton dominates at early eras after the end of inflation but before Big Bang Nucleosynthesis (BBN). The latter ensures that BBN  is not upset by the presence of the dilaton. 
In this paper, within the context of such a scenario, we study the phenomenology of the constrained minimal supersymmetric model (CMSSM) by taking into account all recent experimental constraints, including those from the LHC searches. We find that the allowed parameter space is greatly enlarged and includes regions that are beyond the reach of LHC. 
The allowed regions are compatible with Direct Dark Matter searches since the small neutralino annihilation rates, that are now in accord with the cosmological data on the relic density, imply  small 
neutralino-nucleon  cross sections below the sensitivities of the Direct Dark Matter experiments. 
It is also important that the new cosmologically accepted regions are compatible with Higgs boson masses larger than 120 GeV, as 
it is indicated  from the LHC experimental data. 
The smaller annihilation cross sections needed to explain WMAP data require that the detector performances of  current and planned indirect DM search experiments through $\gamma-$rays should be greatly  improved in order to probe 
the CMSSM regions. 
\end{abstract}
%%%
\maketitle
%\vspace*{8cm}
%%%%%%%%%%%%%%%%%%%%%%%%%% Paper body %%%%%%%%%%%%%%%%%%%%%%%%%
%%%%%%%%%%%%%%%%%%%%%%%%%%%%%%%%%%%%%%%%%%%%%%%%%%%%%%%%%%%%%%%
%\newpage

\newpage
\section{Introduction}

Cosmological data accumulated  from various observations over the past twelve years leave little doubt that Dark Matter (DM) and Dark energy (DE) occupy the major 
portion of the total matter-energy budget of the Universe \cite{snIa,cmb,7yrwmap,bao,lensing}. 
These results follow from best-fit analyses of various astrophysical data to the Standard Cosmological Model ($\Lambda$CDM) which can successfully describe the evolution of our Universe. 
The model is based on a Friedmann-Robertson-Walker (FRW) cosmology, which involves cold DM, at a percentage 23\%, baryonic matter at 4\% and a positive cosmological constant
$\Lambda>0$ which is introduced in an ad-hoc manner in an attempt to describe the vacuum energy density. 

The WMAP data~\cite{cmb,7yrwmap}, with their unprecedented  accuracy, have constrained the predictions of various particle physics models, among these supesymmetric (SUSY) theories  that predict the existence of WIMP particles, the neutralinos,  that would play the role of DM. Besides on-going LHC experiments running at $\, \sqrt{s}= 7 \, \TeV  $, with luminosities that are constantly improving, have
constrained the Standard Model (SM) Higgs mass to a narrow window while they have put new limits on the sparticle masses of SUSY theories \cite{lhc_last}. 
The latest CERN LHC data could be interpreted as a first evidence for the existence of a Higgs  with a mass around 125  GeV~ \cite{higgs_dec}. This may have severe implications for all SUSY models, especially if combined with the WMAP data. 

After one year operation the data provided by the LHC  give us the opportunity to delineate regions of the SUSY parameter space that are also compatible with the cosmological data.
In particular, the leading supersymmetric  DM candidates, the neutralino \cite{neutralino}, still lacking experimental verification, results to a DM abundance which is severely restricted by
cosmic microwave background data and various data from collider experiments, if calculated in the context of the simplest supersymmetry models (minimal supersymmetric model embedded
in minimal supergravity~\cite{msugra}). 
In calculating  neutralino DM abundances it is usually assumed that radiation dominates during DM decoupling.
However, there are various mechanisms originating from string theories, due to dilaton dynamics, which assume that our Universe is not radiation dominated during DM decoupling. These may alter the predictions for the DM abundance in a dramatic way.   

In general the existence of scalar fields in the primordial Universe, that contribute to the energy density, may play a significant role and upset the conventional picture. Among these are quintessence scenarios \cite{quint} that have been invoked in an attempt to explain the vacuum energy, whose existence affects relic abundances since Universe is not radiation dominated during DM decoupling. 
In these models  DM relic density is  enhanced \cite{Salati:2002md}, and in some cases this enhancement can reach  the level of 
$ \sim 10^6 $ or so \cite{ullio}. For a review see for instance \cite{quint2}. In general, modifications of the expansion rate which deviate from the standard cosmological scenarios may have dramatic consequences for the DM relic density \cite{Kamionkowski:1990ni,nonstand}. Inversely, it has been pointed out that the  observed amount of DM puts rather strong constraints on possible modifications of the Universe expansion rate at early eras \cite{drees}. A particular subclass of these models, the tracking quintessence scenario, assumes that the quintessence field is in a kination-dominated phase at early eras \cite{quint3}. In this context the predictions for the gravitino and axino DM are considered in \cite{pallis} while in \cite{pallis2} the predictions for the neutralino DM relic, in the popular supersymmetric schemes, is discussed in the light of the observed $ e^{\pm} $-spectrum by PAMELA \cite{pamela} and Fermi-LAT observations \cite{fermi}. 
Also in other string inspired models with a time-dependent dilaton sources~\cite{elmn}, whose evolution obeys a non-equilibrium string dynamics~\cite{cosmo}, the amount of thermal neutralino relic abundance is diluted by factor of ${\cal{O}} (10)$, relative to that calculated within the $\Lambda$ CDM-minimal supergravity
cosmology and such models seem to survive the stringent tests of LHC~\cite{dutta}. In this case the dilution is due to the appearance of a friction-like term,
on the right-hand side of the appropriate Boltzmann equation. This term plays also a significant role in other considerations studied in  \cite{Mavromatos:2010nk}. 

In this paper we explore the phenomenological consequences of the constrained minimal supersymmetric standard model (CMSSM) model, which we consider as a prototype, in order to study the effects of the dilaton field  assuming its energy prevails over radiation  at early eras. In ref.  \cite{prdlahanas} it was shown that the mechanism for the dilution of DM relic abundances is more general relying on more generic features of the dilaton dynamics which are independent of the non-criticality \cite{noncritical} of the underlying string theory. For definiteness we model the dilaton evolution at early eras to be dictated by  exponential-type potentials, which are rather generic in string theories and arise from loop corrections, occurring also in a wide class of quintessence scenarios, inflation models and supergravity theories. The presence of such a dilaton field, that dominates over radiation long before nucleosynthesis, affects relic abundances in a dramatic way due to its dissipative like coupling to Dark Matter. 
Just before and during BBN era dilaton's  energy is much smaller than radiation energy while dilaton attains a constant value at hadronization prohibiting  such a dissipative coupling to ordinary hadrons. Therefore, dilaton has no effect when hadrons are formed and abundances of ordinary matter are not affected, unlike DM relic abundances which can be smaller by factors as small as  
$\, {\cal{O}}(10^{-2}) \,$, as compared  to the conventional calculations. This allows for smaller annihilation cross sections, in the popular supersymmetric schemes employed in literature, and  enlarges the cosmologically allowed parameter space, which is moved to regions that would be otherwise forbidden, altering the potential of discovering supersymmetry at collider experiments.  
%%%%%%%%%%%%%%%%%%%%%%%
We should remark that in this scenario the cross sections of any processes are not affected by the presence of the dilaton, it is the relic density that it does due to the modification of the Boltzmann equation. As we shall see this causes a dilution, in general, and therefore agreement with WMAP data is obtained for smaller annihilation cross sections. This in turn implies smaller elastic 
neutralino-nucleon cross sections which are of relevance to other experiments  aiming at discovering DM (for reviews see \cite{dmrev}), like direct \cite{direct,xenon100} and indirect 
\cite{pamela,fermi,HESS} DM searches (for a review see \cite{conrad}). 
It is known that direct DM experiments impose constraints on supersymmetric models by excluding part of their parametric space that is close to the focus point region. Therefore a phenomenological study requires that both WMAP bounds and those stemming from direct DM searches are considered. 

In this paper modeling the dilaton evolution at early eras as prescribed in \cite{prdlahanas}, we follow a phenomenological analysis by taking into account the recent accelerator data from LHC, the constraints provided by WMAP7, and DM direct searches. 
The current data from indirect detection of DM through $\gamma$-rays are briefly discussed. 
In this discussion we do not consider energy spectra of fluxes from positrons and anti-protons.
Although these can be conditionally be interpreted as due to annihilation of DM \cite{Cirelli:2008pk,Donato:2008jk,Bergstrom:2009fa} nevertheless 
more conventional explanations exist \cite{Bergstrom:2010gh}. In fact, the former may be explained as emission from pulsars and the latter are compatible with 
standard production mechanisms from Cosmic Ray which impinges the stellar gas.  

%%%%%%%%%%%%%%%%%%%%%%
The paper is organized as follows: \\
In section II we outline the salient features pertinent to the dilaton evolution and its dominance at early eras while in section III we discuss the role of the dilaton field and its effect on the DM relic abundance paving the ground for the phenomenological study of the constrained MSSM  model (CMSSM)  which is done in section IV. 
In section V we discuss the importance of $\gamma$-ray indirect DM searches for this scenario. Our conclusions are presented in section VI.
%%%%%%%%%%%%%%%%%%%%%%%%%%%%%%%%%%%%%%%%%%%%%%%%%%%%%%%%%%%%%%%%%%%%%%%%%%%%

\section{Early Dilaton evolution and its dominance}
Since the main hypothesis is  dominance of the dilaton energy, over that 
of  radiation and matter, one can omit their corresponding contributions to the equations of motion. Therefore, the equations that govern the evolution of the dilaton field take on the form
\begin{eqnarray}
&& \ddot \phi + 3 \, H \, \dot \phi + V^{\,'}(\phi) \;=\; 0 \, ,\nonumber \\
&& 3 \, H^2 \;=\; \dfrac{{\dot \phi}^2}{2} + V(\phi) \, ,\nonumber \\ 
&& 2 \, \dot H \;=\; - \, ( \, \varrho_\phi + p_\phi \,) \;=\; - \, {\dot \phi}^2\, . \label{phieq}
\end{eqnarray}
In these the field $\,\phi  $  is dimensionless and the potential carries dimension $mass^2$. The first of these equations is not independent but is derived from the other two. 
We further assume  that the evolution of the dilaton is dictated by  an exponential-type  potential having the form  $\; V \sim e^{- \, k \, \phi} \; $. Such potentials 
are motivated by inflation \cite{expopot}, 
quintessence scenarios \cite{expoqui} and they can also occur in string theories as perturbative or non-perturbative corrections. Such potentials accept dilaton solutions  which are logarithmically dependent on the cosmic scale factor $\, a(t) \, $, during early eras, having the form 
\begin{eqnarray}
\phi \;=\; c \;  \ln \, \left( \dfrac{a}{a_I} \right) \;+\; \phi_I \, .
\label{phi}
\end{eqnarray}
In this $c$  is a constant and  $ a_I \equiv a(t_I)$  is the cosmic scale factor at the maximal reheating temperature, which was reached after inflation, denoted hereafter by $T_I$. The corresponding time at which this occurred is $t_I $.  

Evidently the above solution for the dilaton holds in epochs $ t_I < t < t_X $ or, in terms of the cosmic scale factor, when $\,a_I > a > a_X  $. Throughout the subscript $X$ signals the end of the period in which dilaton dominates over radiation and matter. At $t_X$ dilaton stops evolving, it attains a constant value so that its kinetic energy vanishes, and its potential energy becomes very small so that radiation energy starts taking over. 

Both $a_I$ and $a_X$ are inputs and it is more convenient to trade them for the parameters  $r$ and $b$ defined by $\,r \equiv \ln(a_I/a_0)$ and 
$\,b \equiv \ln(a_X/a_0)$ respectively.
The first is set by the choice of the reheating temperature $\,T_I$. 
In a supersymmetric model with the MSSM content, whose sparticle mass spectrum is in the range of a few TeV, supersymmetric as well as SM particles are all relativistic and the effective number of degrees of freedom is $\, g_{eff} = 228. 75  \,$, independently of the precise sparticle mass spectrum. This entails to a value $\, r \,= -50.86 -  \ln( \,T_I / 10^{\,9} \GeV \,)  $.
Throughout we assume that $\,T_I$ is of the order of $\, 10^{\,9} \, \GeV$ and hence $r$ is in the vicinity of $ \sim - 50 $.
Although our conclusions are not sensitive to the value of  $\, T_I$, 
as long as it is much higher than the $\TeV$ scale, i.e. the typical supersymmetry breaking scale, in this work for definiteness we consider values of it in the  $\, 10^9 \, \GeV$  range which are supported by CMB constraints, which put lower bounds on the inflationary reheating temperature \cite{Martin:2010kz}, of the order of $\sim 10^6 \, \GeV$,  and Baryogenesis through thermal Leptogenesis  which  demands temperatures  $\sim 10^9 \, \GeV \,$  \cite{Leptogenesis}.
%%%%%%%%%%%%%%%%%%%%%%%

As for the other parameter $\,b$, a first upper bound arises from BBN which should not be affected by the presence of the dilaton. Therefore, the end of the dilaton dominance period $\,t_X$ must occur before Nucleosynthesis. In particular just after $\,t_X$ radiation starts contributing and  overwhelms dilaton's energy at the time of Nucleosynthesis. The latter took place at 
$T_{BBN} \simeq 1 \MeV $, corresponding to $ \ln \, (a_{BBN} / a_0 )\,\simeq -22.5 $ and therefore $\,b$ must be less than this. 
However the bound put on $\,b$  must be even smaller since dilaton should get an almost constant value when hadrons are non-relativistic and this occured before Nucleosynthesis.   
Its constancy is rather mandatory  in this era, otherwise the diluting dilaton mechanism will 
affect the abundances of the known hadrons and especially nucleons which we do not want to occur. 
In fact, the couplings of dilaton  to matter density appear through dissipative terms   
$ \sim \,( \varrho_m \, - 3 \, p_m ) \, \dot{\phi}  $, which modify the corresponding continuity equation for matter.   
Such terms are vanishing when hadrons are relativistic, that is 
at temperatures higher than about $ T_h \sim 1 \, \GeV  $, corresponding to $  \ln \, (a_h / a_0 ) \,\sim -30 $, causing no harm. However below 
$\, T_h  \,  $ hadrons are pressureless  and dilaton couples to hadrons through $ \sim \, \varrho_m \, \dot{\phi}  $. Therefore, in this temperature regime the dilaton has to be almost constant in order to suppress its coupling to hadronic matter. A reasonable value is  $ T_h = \Lambda_{QCD}   $, 
with $ \Lambda_{QCD} \simeq 260 \MeV  $ the characteristic QCD scale. This lowers   $  a_X  $, and hence $b$, suggesting a value given 
by $ b \simeq - 28.4 $. The interesting point is that for such values of $ T_h  $  the coupling of the dilaton to supersymmetric matter is non-vanishing during DM decoupling. In particular, if DM has supersymmetric origin, its decoupling 
occurs at a temperature between  $ T_{DM} \simeq \, 5 -  20 \, \GeV $,  corresponding to  values of the cosmic scale factor  
$  \ln \, (a_{DM} / a_0 )  $ in the range $  \simeq -31.5  $ to $  \simeq  -33.0 $, when dilaton is still the dominant source of energy. 
This may have dramatic effects for the  DM relic abundances as we shall see. 

As a side remark, note that a constant dilaton in the  range $ t > t_X$ cannot account for changes of the fine structure constant,  
$ \Delta \alpha / \alpha \, \sim 10^{-5} $,   over cosmological time scales. The constancy of the dilaton in this era naturally implies small quantum fluctuations $ \Delta \, \phi \ll 1 $ which may put limits on the dilaton-matter coupling   approaching the capability of E{\"{o}}tvos-like experiments.
%%%%%%%%%%%

%%%%%%%%%%%%%%%%%%%%%%%%%%%
A dilaton solution given by Eq. (\ref{phi}) implies that the time derivative of $ \phi $ is linearly related to the Hubble rate 
through  $ \dot \phi \,= \, c \, H  \,$,  for  $ t < t_X  $. 
Then by the third of equations (\ref{phieq}), which can be solved for  $ H$, we find that $ H $ is inverse proportional of the time $ t $.  
Solving   $ H = \dot a / a   $ we can get the expansion rate $\,a(t) \,$  and from this the form of the dilaton field as function of the time,   
\begin{eqnarray}
\phi \;=\; \dfrac{2}{c} \, \ln \, {\left( \dfrac{c^2 \, H_I}{2} \, t_*  + 1 \right)} + \phi_I  \, .
\label{phit}
\end{eqnarray}
%%%%%%%%%%%
In this  subscripts I denote quantities evaluated at $\; t_I \;$ and $\, t_* \equiv t - t_I $. 
%%% From this and using Eq. (\ref{phi}) we can derive the expansion rate $\,a(t)$ as function of time. 

Knowing the dilaton and the Hubble rate, from the second of Eqs. (\ref{phieq}) the dependence of the potential on the values of the expansion rate and the dilaton at the reheating time can be derived
\begin{eqnarray}
V(\phi) \;=\; 
\left( \frac{6 - c^2}{2}  \right) \; H_I^{\,2} \; e^{ - \, c \, ( \phi - \phi_I ) } \, .
\label{pot}
\end{eqnarray}
%%%%%%%%%%%%%%%%
Using this one finds that the value of the potential at $a_X$ is exponentially suppressed relative to that at $a_I$. In fact  
\begin{eqnarray}
V(\phi_X ) \, / \,  V(\phi_I ) \,=\, \exp \,[ \,{-c \, ( \phi_X-\phi_I )}\, ] \sim \exp \,(\,{- B \,c^2 \,) }  \, .
\label{exxx}
\end{eqnarray}
In this the parameter $\,B$ is $\, B = b -r $, with $b, \, r$ defined before,  and  therefore $\,B$ has a value around  $\sim 20$. 
For a positive potential, which drops as Universe expands, the constant $c$ should be bounded by $\, c^2 < 6  \,$,  
as is evident from (\ref{pot}), while  as we shall see, dominance of the dilaton energy over radiation is achieved for $\,c^2 > 4 \,$.  Since $\,B$ has a value 
$\simeq 20$, by Eq. (\ref{exxx}) we see that  dilaton's potential energy is exponentially suppressed long before Nucleosynthesis.  
In fact, the ratio of the potential energy, at the end of the dilaton-dominated era, to the same energy at reheating temperature, given above, is found to drop by at least fourty orders of magnitude in this scenario, and the same holds for the total dilaton's energy since the ratio of its kinetic to its potential energy is constant. This is easily derived by using the second of Eq. (\ref{phieq}),
using the fact that  $\, \dot \phi = c \, H \,  $. Hence in this model the barotropic index $w$ has a constant value, in the regime $\, t_I < t < t_X \,  $, given by $\, w = c^2/3 -1  \,$.  From the limits put on $\,c^2\,$ we  conclude that w lies in the range  $\, 1 < w < 2 \,$.  
%%%%%%%%%%%%%%%%%%%%%%

For the dilaton to radiation energy density ratio one finds that
%%%%%%%%%%%%%%%
\begin{eqnarray}
\frac{  \hat{\rho}_{\,\phi}  }{  \hat{\rho}_{\,r} }
\,=\, {\left( \frac{  \hat{\rho}_{\,\phi}  }{  \hat{\rho}_{\,r} } \right)} \bigg{\arrowvert}_I \; 
 {\left( \frac{a}{a_I} \right)}^{\,4 - c^2 } \, ,
\label{ratior3}
\end{eqnarray}
%%%%%%%%%%%
where $\, { (  \hat{\rho}_{\,\phi}  }/{  \hat{\rho}_{\,r} ) } {\arrowvert}_I \,$ is the same ratio at the reheating temperature  $\, T_I \,$. 
In Eq. (\ref{ratior3}) we have reinstated dimensions and hatted densities carry  dimension ${\mathrm{energy}}^{4 }$. This ratio ought to be much larger than unity in the entire range 
$\, t_I < t < t_X \,$, since dilaton dominates over radiation in this period, and should decrease as  Universe expands so that eventually  before Nucleosynthesis radiation  starts dominating.  For this to happen the constant $\, c$ should satisfy 
$\, c^2 > 4 \,$. Combined with the bound $ \, c^2 < 6\, $ discussed earlier, which ensures that we have a positive potential that drops as Universe cools, we see that $\,c\,$ should lie in the rather tight range  $ \, 4 < c^2 < 6\, $. 
As a remark we point out that the ratio $\, { (  \hat{\rho}_{\,\phi}  }/{  \hat{\rho}_{\,r} ) } {\arrowvert}_I \,$  turns out to be 
 proportional to $\, {(\, H_I / H_0 \,) }^2 $, where $\, H_I $ is the value of the Hubble rate at the end of inflation and $\, H_0 $ its corresponding value today. 
Measurements of the power spectrum of scalar and tensor perturbations yield bounds on the inflationary potential which in turn
 imply an upper bound for $\, {(\, H_I / H_0 \,) } $, \cite{Smith:2005mm}. However this bound is rather weak in imposing a severe upper bound  on 
the ratio $\, { (  \hat{\rho}_{\,\phi}  }/{  \hat{\rho}_{\,r} ) } {\arrowvert}_I \,$ which can be comfortably much larger than unity, as demanded,  conforming at the same time with the constraints arising from the power spectrum measurements (for details see \cite{prdlahanas}).

%%%%%%%%%%%
\section{Dilution of Relic Densities}

Due to the presence of the dissipative term the continuity equation for the energy-matter density includes an additional term, 
\begin{eqnarray}
\frac{d \rho}{dt}
 +3 \hat{H} (\;{\rho} + p ) -\dfrac{\dot \phi}{\sqrt{2}  } \; ({\rho} - 3 p )=0 \;   .
 \label{continuity}
\end{eqnarray}
%%%%%%%%
In this equation the division by $ \sqrt{2} $ of the last term is due to the normalization 
of the dilaton whose kinetic energy in our notation appears as  $ \rho_\phi^{kin} = {{\dot \phi}^2} / {2} $ as is  seen from  the second of Eq. (\ref{phieq}).
It should be noted that throughout this work we are assuming  the lowest order, in $\alpha^\prime$, contributions to the form factors 
$\; e^{-\psi(\phi)} \;$ and $\; Z(\phi) \;$ associated with the scalar curvature $R$ and dilaton kinetic terms of the string effective action, in the string frame, and hence the simple expressions for the $\dot{\phi} $-dependent term of Eq. (\ref{continuity}). Also for simplicity the dilatonic charge has been taken vanishing. We are aware of the fact that the couplings of the dilaton to matter may evolve in time with the dilaton itself and depend on the particle species in a non-universal way. Therefore, other options are available \cite{Gasperini:2001mr,Gasperini:2001pc}. Besides there may also exist additional contributions to the continuity equation which depend on the dilatonic charges of the species under consideration as studied in 
\cite{Gasperini:2001mr,Gasperini:2001pc,Gasperini:2002bn}. Certainly in order to take into account all these effects one needs a better understanding of the underlying string dynamics including perturbative and non-perturbative string effects which are lacking. Therefore, in this work we consider the simple case scenario in which the gravi-dilaton effective action is approximated by its lowest order Lagrangian in the string slope $\alpha^\prime$. 

The last term  in Eq.~(\ref{continuity})  drops  when matter is relativistic.  Obviously this is also the case for radiation since  the difference  $\,{\rho} - 3 p  $   stays zero during the entire cosmological evolution. Therefore, for particles of a certain mass $\,m$ the last term drops for temperatures $\, T \gg m \,$ and the effect of the dilaton dissipative force is absent. However, as temperature drops this term eventually  starts contributing for all period until the dilaton attains its constancy which occurs after DM decoupling but before hadron formation, and thus before BBN, in the way described in the previous section. Including the collision term and neglecting the pressure, for temperatures in the regime $\, T <  m \,$,  Eq. (\ref{continuity}) implies  the following equation for the number density 
%%%%%%%%%%%%%%%%%%%%%%%%%%%%%%
\begin{eqnarray}
\frac{dn}{dt} + 3 H n + \vev{  \sigma  \,  v} (n^2 -n^2_{eq}) - \dfrac{\dot \phi}{\sqrt{2}  }\;n \;=\;0 \, .
\label{dndt}
\end{eqnarray}
%%%%%%%%%%%%%%%%%%%%%%%%%%%%
This is suitable for describing the evolution of the number density during periods for which the particle is non-relativistic and pressure practically vanishes. During eras in which the particle is relativistic the last term in Eq. (\ref{dndt}) is absent, as we argued before, since the last term in the continuity equation  (\ref{continuity}) is absent too. Thus  Eq.~(\ref{dndt}) receives the well-known form of Boltzmann equation \cite{boltz}. 
Using the number to entropy density ratio, $\, Y \,=\,n / s \,$, this equation takes on the following form  
%%%%%%%%%%%%%%
\bear
\frac{dY}{dx}=\xi(x) \; \; m \vev{ \sigma  \,  v}\; \left(  \frac{45\,G_N}{\pi} {g}_{eff} \right)^{-1/2} \;( h+ \frac{x}{3} \frac{dh}{dx})\; ( Y^2-Y^2_{eq}) \;+\; S(x) \; \; Y  \, ,
%%- \frac{\Gamma}{H x} \; ( 1+ \frac{x}{3 h} \frac{dh}{dx}) \; Y \quad 
\label{boly}
\eear
%%%%%%%%
%%%%%%%%
where $x$ stands for $\, x = T / m  \,$, with   $T$ the photon gas temperature which is related to the radiation density $\rho_r \; $  through
$$
\rho_r \;=\; \frac{\pi^2}{30} \; g_{eff}(T) \; T^4 \, ,
$$
%%%%%%%%%%%%%%%%%%%%%%%%%%%%%%%%%
The function  $\,g_{eff}(T)  $, at a given epoch, counts the degrees of freedom of the particles that are relativistic that epoch. 
In Eq. (\ref{boly}) $G_N $ is Newton's constant and  the quantity $h$ stands for the entropic degrees of freedom related to the entropy density through 
$\, s = 2 \pi^2  \, T^3 \,h(T) \, / \, 45   \,$. Note  the appearance of the  the prefactor $\, \xi(x) \,$ in the first term of  Eq (\ref{boly}) which is given by 
%%%%%%%%%%%%%
\bear
\xi(x) \;=\; {\left( 1 + \frac{\rho_m}{\rho_{\,r}} +  \frac{\rho_\phi}{8 \pi G_N \, \rho_{\,r}}    \right)}^{-1/2} \, .
\label{xifactor}
\eear
%%%%%%%%%%%%%%%%%%%%
In this 
$\, \rho_{\,r} , \rho_{\,m}  \,$ and $\,  \rho_\phi  \,$ are the radiation, matter and dilaton energy-densities respectively. 
Recall that we use a dilaton density having dimensions $ \, m_p^2 \, $, see Eq. \ref{phieq}. Note also that no cosmological-term contributes to $\, \xi(x) \,$ since such a term is absent during the early Universe evolution which includes the DM decoupling era. $\, \xi(x) \,$ is smaller than unity having  therefore the tendency to decrease relic abundances at slower rates as compared to the conventional cases, where this factor is unity due to the fact that in these  approaches Universe is assumed radiation dominated during DM decoupling. 
Therefore, if it were only for $\, \xi(x) \,$  an increase of the DM relic density, relative to the conventional scenarios, would have been predicted. However in the same equation 
an additional source term $\, S(x) \,$ is present, given by 
%%%%%%%%%%%%%
\bear
S(x) \;=\; - \frac{\; \, \phi^{\, \prime}}{\sqrt{2} \,x} \; \left( 1+ \frac{x}{3 h} \, \frac{dh}{dx}  \right) \, .
\label{source}
\eear
%%%%%%%%%%%%%%%%%%%%
In this expression $\,\phi^{\, \prime}  $  denotes differentiation with respect $\ln(a/a_0)$. If dilaton decreases, with decreasing temperature, then 
the source term is positive and acts in the opposite direction tending to decrease $\,Y$ after DM decoupling. Therefore, two different mechanisms compete in this scenario and the net effect may be a decrease of the relic abundance. 

%\newpage
%%%%%%%%%%%
\begin{figure}
\begin{center}
\includegraphics[width=7.5cm]{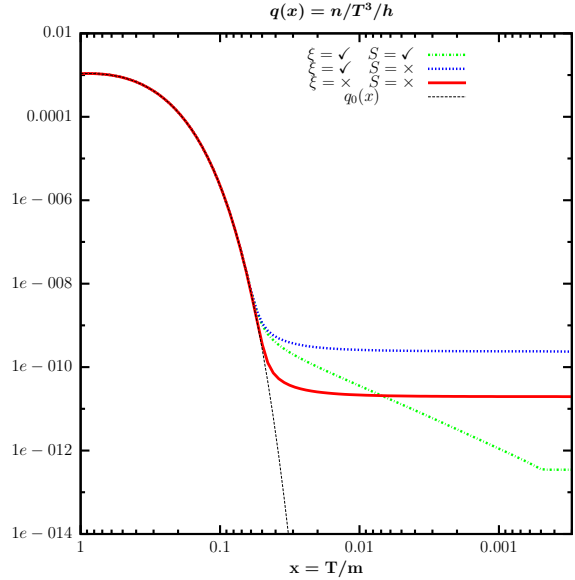}
\end{center}
\caption[]{
The LSP Dark Matter number density to entropy density ratio $  {q = \frac{n}{T^3 h} } $ as function of 
$\, x = \frac{T }{m_\mathrm{LSP}} \, $ in a particular supergavity model. The values of $ \xi , S $ denote the status of the $\xi$-factor and the source respectively  ( $\checkmark $ for open, $\times $ for switched-off ). For comparison the corresponding equilibrium density $ q_0(x) $ has been also drawn. 
(From Ref.~\cite{prdlahanas}.)
}
\label{fig0}  
\end{figure}
%%%%%%%%%%%%%%%%%%%%%%%%%%

%%%%%%%%%%%%%%%%%%%%%%%%%%%%%%%%%%%%%%%%%%%%%%%%
As an example, 
in Fig.~\ref{fig0}  we display the effects of the 
the presence of the factors $ \, \xi(x)$ and $\, S(x) $, as given by Eqs (\ref{xifactor}) and 
(\ref{source}), for the ratio of the number to entropy density  $ n /( T^3 h )  $. 
The displayed figure correspond to a supergravity model with inputs given by $\, m_0=1100 \GeV ,\, M_{1/2} =1200 \GeV  \,$ and $\, A_0 = 0 \GeV  \,$
 with  $\tan \beta = 40 $ and the parameter 
$\mu > 0 $. For the particular SUSY inputs the  LSP Bino has a mass $m_\mathrm{LSP} = 527.2 \GeV  $. 
Eq (\ref{boly}) has been integrated  numerically which yields more reliable  results than other approximate schemes employed in other works. 
 The value of  $\,b \equiv \ln(a_X/a_0) \,$, setting the onset of the epoch after which the dilaton is constant, has been taken  $ - 28.4 $ corresponding to a temperature $\,\Lambda_{QCD} = 260 \GeV$ as we have already discussed. For comparison in Fig.~\ref{fig0} except the ordinary case scenario, where both the source $S$ and the 
$\xi - $ factor are absent (red solid line), 
the cases where both $\xi $  and $S$ are open (green dashed-dotted line), or when only the $\xi - $ factor is present (blue short-dashed line) are also shown. The very thin dashed line, that rapidly drops, is the equilibrium density. In the case, that both terms are switched on, the density is monotonically decreasing after decoupling due to the appearance of the source term. The rapid change around 
$ x \simeq 0.0005 $, corresponds to a value of the cosmic scale factor specified by $b = - 28.4 $ where dilaton reaches its  constancy. In the specific example shown the relic density is diluted by a factor of $\sim 50  $, as can be seen by comparing today's density predictions for the conventional case (red solid line) and the case where both $\xi - $ factor and the source term are switched on (green dashed line). In the first case the relic density predicted 
is $ \Omega_\mathrm{LSP} \, h_0^2 \, = \, 6.059 \, $ while in the second case the relic density is considerably reduced falling into the WMAP allowed range $ \, \Omega_\mathrm{LSP} \, h_0^2 \, = \, 0.1116 \, $. 
%%%%%%%%%%%%%%%%%%%%%%%%%%%%%%%%%%%%%%%%%%%%%%%%%%%%%%%%%%%%%%%%%%%%%%%%%

Although in this work  we solve (\ref{boly}) numerically, which yields the most accurate and reliable results, we shall proceed to a semi-analytic treatment of (\ref{boly}) and illuminate some of its qualitative features that are worth pointing out.  
The presence of $\, \xi(x)$ in this equation it effectively decreases the thermal average $\, \vev{  \sigma  \,  v}$ since $\, \xi(x) < 1  $. As a consequence this term has the tendency to shift the freeze-out point $\, x_f = T_f / m $  to higher values, as compared  to the conventional case where this term is unity, while the source term has little effect on $\, x_f  $.  We shall quantify this later. This results to a larger value of $\, Y$ at $\,x_f$. Shortly after freeze-out $\, \vev{  \sigma v} \, n \ll H $ and the first term in (\ref{boly}) can be dropped. Then we can integrate from $\, x_f$ to today's temperature and find 
\begin{eqnarray}
\ln \,  \frac{Y(x_f)}{Y(x)}    \,= \, \int_x^{x_f} \, S(x) \, dx  \quad .
\label{fromf}
\end{eqnarray}
Integrations of the source   $\,S(x)$ can be performed analytically,  
if the entropy contributions in (\ref{source}) are neglected,  which is a decent approximation. In particular
\begin{eqnarray}
\int_{x_1}^{{x}_{2}} \, S(x)\, dx \;=\; \frac{\phi(x_2) - \phi(x_1)}{\sqrt{2}} \; = \frac{c \; \ln \, (a_2/a_1)}{\sqrt{2}} \,  \;=\;
\frac{c}{\sqrt{2}} \, \left( \,  \ln \, \frac{T_1}{T_2} 
+ \frac{1}{4} \, \ln \, \frac{ g_{eff}(T_1)}{  g_{eff}(T_2) }    \right)   \, ,
\label{integrations}
\end{eqnarray}
where the particular dilaton solution (\ref{phi}) has been employed. If the lower limit $\, x_1$ of the integration is below $\, x_X \equiv T_X / m  $   then $\, T_1$ should be replaced by  $\,T_X$ on the r.h.s.
Therefore from (\ref{fromf}) and (\ref{integrations}) we get 
\begin{eqnarray}
\ln Y(x) \; = \; \ln Y(x_f) - 
\frac{c}{\sqrt{2}} \, \left( \,  \ln \, \frac{x}{x_f} 
+ \frac{1}{4} \, \ln \, \frac{ g_{eff}(x)}{  g_{eff}(x_f) } \right)  \, ,
\label{solution}
\end{eqnarray}
for any $\, x < x_f$. Since $\, c < 0$ the logarithmic term $\, \sim \ln x$ explains the linear drop  of $\, \ln Y(x)$ below $x_f$ in  Fig.~\ref{fig0}  (green line). 
From this it can be seen that 
the value of $Y$ reached at today's temperature is rather sensitive to the choice of the parameter $\, b$. In particular for two different values $b$ and $b^\prime$ the resulting values are related by
\begin{eqnarray}
 Y^\prime(x_0) \; = \;  Y(x_0) \; \exp   \frac{c \,  ( b^\prime - b ) }{\sqrt{2} }   \, .
\label{solution2}
\end{eqnarray}
As the value of $\, c $ is confined in the range $\, 2 < | c | < 2.45 $ we observe that by increasing $\,b$  by a unit,  
$\, Y_0^\prime$  decreases relative to $\, Y_0$ by factor of about 
of $\simeq 5 $. 
%%%%%%%%%%%%%%%%%%%%%%%%%%%%%%%%%%%%%%%%%%%%%%%%%%%%%%%%%%%%%%%%%%%%%%%%%%%%

One can attempt to find semi-analytic solutions of Eq. (\ref{boly}) which constitute approximations of the numerical results. 
Using standard techniques, already encountered in the dilaton-free case, where the terms $\, \xi , S$     are absent one can solve (\ref{boly}) in a semi analytic way. This treatment leads to the following result for the relic density 
{\footnote{Such a treatment has been already considered in the first of \cite{elmn}. 
Here we refine and modify the results presented in that work and adapt them to the model at hand.}
\begin{eqnarray}
\Omega \, h_0^2 \;=\; R \;  { ( \Omega \, h_0^2 )}_{no-dilaton}    \, ,
\label{omega}
\end{eqnarray}
where the factor $\,R$ on the r.h.s. accounts for a change of the result of the conventional case. In this approximation $\,R$ is given by 
\begin{eqnarray}
R \; = \; k \,  \xi^{-1}(x_f) \;  \exp{\left[ -  \, \int_{x_0}^{{x}_{f}} \, S(x)\, dx \right]  }   \, ,
\label{rfactor}
\end{eqnarray}
where $\,k$ is a constant, with $\, k < 1$, which  will  be discussed below. 
Wherever appears $\,x_f$    stands for the freeze-out temperature   and  $\, x_0$  is the  value of $\, x = T / m $ today. 
The integrations of the source   $\,S(x)$ in this equation can be performed analytically, as we have already discussed, using  
 (\ref{integrations}). 

In the approximate scheme that leads to (\ref{omega}) the freeze-out point is found by solving  
\begin{eqnarray}
x_f^{-1} \;=\; 
&& \ln \left(  0.076 \, c_f (c_f+2) \, m \, M_\mathrm{Planck}  \, x_f^{1/2}  \, \vev{  \sigma  \,  v}_f \, g_{ eff}^{- 1/2}(x_f)  \right)  \nonumber \\
 && + \ln \, \xi(x_f) - 
 \ln \left(1 - 3 x_f / 2 - x_f^2 \, (1+c_f) \, S(x_f)  \right)    \, ,
\label{freezeit}
\end{eqnarray}
where $\, M_\mathrm{Planck} = 1.22 \times 10^{19} \,$ GeV. 
When $\, \xi = 1$ and $\, S=0$ the constant $\,k$ becomes unity and we recover the standard result. 
In this equation the constant $\, c_f$ is of order unity  and it is defined by $\, Y - Y_{eq} = c_f Y_{eq}$. In the standard cases the value of 
$\, c_f$ is adjusted so that agreement with the approximate results is obtained. 
The source term at $\, x_f$ is $\, S(x_f) =  - c \, /  \sqrt{2} \, x_f $ and thus the last term in (\ref{freezeit}) can be shown to be a small number if $\; x_f$ is $\, {\cal{ O}}( 0.01)$. The quantity  $\, \xi(x)  $ is the inverse square root of the ratio of the dilaton to radiation energy at $\,x$, when dilaton energy dominates. Therefore  
the $\ln \, \xi(x_f)  $  term in (\ref{freezeit}) is negative and it dominates over the last term in the same equation.  Hence, the freeze out point is shifted towards larger values as compared to the standard result. 
For instance, with values of the dilaton to radiation energy ratio  $\, \sim 10^2 - 10^4$, in the region $ T=5-50 \, \GeV  $, one can find from  (\ref{freezeit}) that the relevant shift is at most $ 20 \%$. This we have also verified numerically. 
Therefore for a WIMP, values in the range $ \sim 0.05-06 $ are expected for the freeze-out point, provided the ratio of the dilaton to radiation energy is not exceedingly large.    

Regarding the factor $\, R$ of Eq. (\ref{rfactor}), the factor $\,  \xi^{-1}(x_f) $,  
and the constant  $\,k$, arose by approximating the integral $\, \int_0^{x_f} \xi(x) \, \vev{  \sigma  \,  v }  dx $, 
encountered in integrating   (\ref{boly}) from $\, x_0$ to $\, x_f$, by 
$\, \xi(x_f) \, \int_0^{x_f} \, \vev{  \sigma  \,  v }  dx   $. 
The constant $\, k$ is actually the ratio of the approximated to the exact integral. 
Since $\, \xi(x) > \xi(x_f)$, for any $ \, x < x_f $, by its definition the constant $\, k$ is less than unity. 
Our numerical procedure shows that $\, k$ takes values in the whole region  from $\, 0.1 $ to $\, 1.0 $,  with no preference to particular values or region,  and  
it depends on the  inputs in  a  way that it does not allow us to express  it by a simple empirical formula. Hence $\, R$ as it stands cannot be trusted to derive precise results 
and this is the main reason we decide to solve Boltzmann equation numerically. However $\, R$ bears the main qualitative features, especially the exponential factor suppressing the relic density and it serves as an order of magnitude estimate of the actual result. 
In $R$, the factor $\, \xi^{-1}(x_f)$  tends  to increase relic density, as is evident from Eq. (\ref{omega}). 
The enhancement caused by $\, \xi^{-1}(x_f)$  is followed by the exponential term in Eq. (\ref{rfactor}) that acts in the opposite direction tending to decrease the relic density.

$\, R$  depends on $\, T_I$,  $\, T_X$, which sets the end of the dilaton dominance era, and the value of the slope $\,c$, which are  free parameters. Besides the ratio of the dilaton to radiation energy density   $\, \Delta = {\rho_\phi}/{8 \pi G_N \, \rho_{\,r}} $ at a given temperature, say $\, T_X$, which is shall denote by $\, \Delta_X $,  is also a free parameter. Given $\, \Delta_X $, by Eq. 
(\ref{ratior3}), its value at any other temperature can be derived. 
Although $\, T_I$ is an arbitrary parameter the preferred values of it are around $10^9 \, \GeV$ as dictated by inflation  and Leptogenesis scenarios as we have already pointed out. 
The constant $\, c$ lies in the narrow range $\,2< |c| < 2.45$, as we have already stated and $\,T_X$ should take values larger than the typical 
hadronic scale, $T_h \simeq 1\, \GeV$ or less, but not much less. In our numerical studies we use values  $\, T_X \simeq \Lambda_{QCD} $ which yield the maximal suppression to the relic density.  

It facilitates  discussion  if $\, R$ is cast in the following form 
\begin{eqnarray}
R \;=\; k \, e^{- \, P} \quad .
\label{PPP}
\end{eqnarray}
In this the exponent $\,P$, to a  good approximation, is given by 
\begin{eqnarray}
P \;=\; \frac{1}{2} \, ( \, 4 - c^2 - \sqrt{2} \, c \,) \; \ln \, \frac{T_f}{T_X} \; - \, \frac{1}{2} \, \ln \, 
( 1 + \Delta_X )   \, .
\label{PPP2}
\end{eqnarray}
From this we see that suppression of the relic density is obtained if $\, P>0$. In $\,P$ the prefactor of $\, \ln \, \frac{T_f}{T_X} $ is positive in the allowed range $\,2< |c| < 2.45$, taking values varying from $ 1.41 $ to $0.73$. Since $T_f > T_X$ the first term suppresses $\,R$, and hence relic density, while the second term tends to increase it. 
The maximal suppression effect is implemented for the smallest possible values of $\, |c|$,  $T_X$ and $\,\Delta_X  $. As a  case leading to a severe reduction of the relic density, if we take for instance $\,T_X \simeq 0.5 \, \GeV$, $\, T_f \simeq 20 \, \GeV$ $ \,|c|=  2.05 $ and $\, \Delta_X \simeq 100 $  from the exponential term alone we get $\, R \simeq  0.065  $ which is of order $\, {\cal{O}}(10^{-2}) $. The prefactor $\,k$ decreases further this value. 
%%%%%%%%%%%%%%%%%%%%%%%%%%%%%%%%%%%%%%%%%%%%%%%%%%%%%%%%%%%%%%%%%%%%%%%%%%%%%%
\begin{table}
%\hspace*{-14mm}
%\begin{center}
%\small
\addtolength{\tabcolsep}{2pt}
%%%\scalebox{0.85}{%
\begin{tabular}{cc|c|ccccc} 
%\hline 
%\multicolumn{2}{c|}{ a \quad \quad b} & \multicolumn{1}{c|}{ $ m_{\tilde{\chi} } [\, GeV \,] $ } 
\hline
{\rule{0pt}{4.4ex}}
a        &  b     & \quad $ \; m_{{\chi} }  $  \quad \quad  & \quad  \quad $ x_f^c $ \quad  \quad&   $ 10^{\,2} \times R $  &   $ \Omega h_0^2 \, \arrowvert_{num}  $  &  $ \Omega h_0^2\, \arrowvert_{app}  $   & $ \dfrac{\Omega h_0^2 \, \arrowvert_{num} }{ \Omega h_0^2\, \arrowvert_{app} }$ 
\\  % [12pt]
\multicolumn{2}{c|}{  $ \; [\, \mathrm{cm^3/s} \,] $} & \multicolumn{1}{c|}{ $  [\, \GeV \,] $ } &  \multicolumn{5}{c}{  } \\ [4pt] \hline \hline
{\rule{0pt}{2.5ex}}
                             &           &      500    &        0.0329    &    2.987     &     4.741 $\times 10^{-2}$     &    4.062 $\times 10^{-2}$    &    1.167 \\
2.5 $\times$ $10^{-25} $     &       0   &      900    &        0.0322    &    1.343     &     2.154 $\times 10^{-2}$     &    1.822 $\times 10^{-2}$    &    1.182 \\
                             &           &      5000   &        0.0304    &    0.129     &     0.0020 $\times 10^{-2}$     &    0.0017 $\times 10^{-2}$    &    1.176 \\ [2pt]
\hline
{\rule{0pt}{2.5ex}}
                             &          &      250    &        0.0441    &    5.240     &     0.1542     &    0.1346    &    1.146 \\
1.0 $\times$ $10^{-27} $     &       0  &      800    &        0.0418    &    1.096     &     3.380 $\times 10^{-2}$    &    2.859 $\times 10^{-2}$    &    1.182 \\
                             &          &      5000   &        0.0387    &    0.092     &     0.273 $\times 10^{-2}$    &    0.230 $\times 10^{-2}$    &    1.187 \\ [2pt]
\hline
{\rule{0pt}{2.5ex}}
                             &          &      250    &        0.0592    &    3.465     &     7.656     &    6.678  &    1.146 \\
1.0 $\times$ $10^{-29} $     &       0  &      800    &        0.0552    &    0.741     &     1.711     &    1.453  &    1.178 \\
                             &          &      5000   &        0.0500    &    0.064     &     0.143     &    0.122  &    1.172 \\ [2pt]
\hline
{\rule{0pt}{2.5ex}}
                             &          &      250    &        0.0589    &    3.490     &     7.345    &    6.574   &    1.117 \\
1.0 $\times$ $10^{-29} $     &       1.0 $\times$ $10^{-29} $   &      800    &        0.0549    &    0.746     &     1.674    &    1.432   &    1.170 \\
                             &          &      5000   &        0.0498    &    0.064     &     0.138    &    0.119   &    1.159 \\ [2pt]
\hline
\end{tabular}
%%%%}
\caption{
{\normalsize{
In the first two columns are the values of the coefficients $ \, a , b  $, in $ \mathrm{cm^3 / s}  $, for the thermal averaged cross-section 
$ \vev{  \sigma  \,  v} = a + b \, x    $ while in the third column is the 
WIMP mass. In the next four  columns we display  
the freezing point ( shifted value ) $ x_f^c $, the reduction factor $ 10^2 \times R $,  the relic density derived by solving Boltzmann equation numerically, $ {\Omega h_0^2 \, \arrowvert_{num} } $, and its approximate value   
$ \Omega h_0^2\, \arrowvert_{app}  $.  In the last column their corresponding ratio  $ {\Omega h_0^2 \, \arrowvert_{num} } \, / \, { \Omega h_0^2\, \arrowvert_{app} }$  is shown. 
}}
}
\label{tablecharter}
%\end{center}
\end{table}
%%%%%%%%%%%%%%%%%%%%%
%%%%%%%%%%%%%%%%%%%%%

In the discussion that follows we shall present results for the relic density for  a WIMP particle  of mass   
$\, m_{{\chi} } $ derived by solving Boltzmann equation numerically, and compare it to that obtained by using the semi-analytic method based on Eqs. (\ref{omega},\ref{rfactor}). Although complete agreement between numerical and approximate results cannot be obtained, the comparison will serve in order to check the real magnitude of the reduction and compare it to that implemented by the factor $R$, which was discussed before.  Furthermore such a comparison  may suggest possible refinements which will improve the approximation bridging the discrepancy between approximate and numerical results. 
%%%%%%%%%%%%%

The factor $R$ is dominated by the exponent $P$, given by Eq. (\ref{PPP2}), which however is approximate. 
Its exact form, as read from Eq. (\ref{rfactor}), is 
\begin{eqnarray}
P_{exact} &=& - \frac{1}{2} \ln \, ( 1 + \Delta_f ) +   \, \int_{x_0}^{{x}_{f}} \, S(x)\, dx \nonumber \\
&=& - \frac{1}{2} \,  \ln \left[ 1+ \Delta_X  \, \exp \,(c^2-4) \left( b+ \ln \, \frac{T_f}{T_0} + \frac{1}{4} \, \ln \, \frac{g_{eff}(T_f)}{g_{eff}(T_0)} \right)  \right] \nonumber \\
&& - \frac{ c }{ \sqrt{2}} \, \left( b+ \ln \, \frac{T_f}{T_0} + \frac{1}{4} \, \ln \, \frac{g_{eff}(T_X)}{g_{eff}(T_0)} \right) \,.
\label{longp}
\end{eqnarray}
In passing to the second equation the ratio of the dilaton to radiation energy at $T_f$, denoted by $\Delta_f$, has been expressed in terms of the corresponding ratio at $T_X$, named $\Delta_X$, using 
Eq. (\ref{ratior3}). Also $T_X$ has been expressed in terms of the parameter $b \equiv \ln ( a_I/a_0)$. 
In this way  $P_{exact}$ is expressed in terms of the parameters $b$ , $c$ and $\Delta_X$. 
The freeze-out temperature $T_f$ appearing in Eq. (\ref{longp}) is read from (\ref{freezeit}). The temperature $T_X$ appearing in the last line of this equation  cannot be expressed in an analytic manner in terms of $b$, since $ b = \ln \, (T_0 / T_X) +  {1}/{4} \; \ln ( {g_{eff}(T_X)}/{g_{eff}(T_0)} )$. However for $T_X$ in the range 
from $\Lambda_{QCD}$ to $1 \GeV$, which we are mainly interested in, the quantity 
${1}/{4} \; \ln ( g_{eff}(T_X) / g_{eff}(T_0) ) $  varies little, taking values between $0.66 $ and $0.77 $,
and one can replace it by its average $\simeq 0.71$ reproducing satisfactorily the exact result. In order to fully quantify the factor $R$, and present an expression for it in terms of the  parameters describing the model, the prefactor $k$ appearing in Eq. (\ref{rfactor}) is also needed. This has been discussed before, in this section, and cannot be expressed in an analytic way. Later we shall give an approximate formula which reproduces fairly well the results derived by our numerical procedure. 

%%%%%%%%%%%%
In the following, for simplicity we  are  assuming   a  thermal average for the cross-section times Moeller velocity, given by 
$ \vev{  \sigma  \,  v} \,=\, a + b \, x$, in which S-wave and/or  P-wave annihilation processes are important. In all cases discussed the value of the 
constant $\, c_f$ in Eq. (\ref{freezeit}) is taken $\, c_f = \sqrt{2} - 1$. When  dilaton effects  are absent this value gives the best fit, better than $5 \%$, matching very well numerical and approximate results. In the presence of the dilaton,  better convergence  of (\ref{omega}) to the numerical result is obtained if we use the freeze-out point $\, x_f^c$ 
defined by the criterion  $\, n \vev{  \sigma  \,  v} = H$. This is actually implicit in the calculation that led to 
Eq. (\ref{omega}). For values of  $\, c_f$ that are of order unity  we have checked numerically that $\,x_f^c$ is smaller than $x_f$, derived from Eq.(\ref{freezeit}), having  a value $\, x_f^c = x_f / s $ where the shift coefficient $s$ is close to $s = 1.3$. Using  the shifted value $x_f^c$,  instead of $x_f $,  in  Eqs. (\ref{rfactor}) yields results for the relic density that  approximate better  the numerical values for the relic density as we shall see. Note that the use of $x_f^c$ in (\ref{rfactor}), it  enhances  $\,R$  by a factor of about two not spoiling therefore  the exponential  suppression caused by this factor to the relic density.

Sample results are presented in Table \ref{tablecharter}  where in the first two columns  we give the values of the coefficients $ \, a , b  $,
 in $ \mathrm{cm^3 / s }$, pertinent to the thermal averaged cross-section, and in the third column the WIMP mass. In the next four  columns we display in the following order,  
the "shifted" freezing point, $ x_f^c $, the reduction factor $ 10^2 \times R $,  the relic density as this is derived by solving numerically Boltzmann equation, 
$ {\Omega h_0^2 \, \arrowvert_{num} } $, and its approximate value   $ \Omega h_0^2\, \arrowvert_{app}  $.  In the last column the ratio  
$ {\Omega h_0^2 \, \arrowvert_{num} } \, / \, { \Omega h_0^2\, \arrowvert_{app} }$  is shown. We warn the reader that 
the values of the relic densities appearing in the Table are only for demonstration and are not within the WMAP range. 
In deriving  $ \Omega h_0^2\, \arrowvert_{app}  $, the factor $k$ appearing in  Eq. ( \ref{rfactor}) has been  approximated by taking the  average value     
$\, \xi(x) = ( \xi(x_f) + \xi(x_X)  ) /2 $ within the integral  $\, \int_0^{x_f} \xi(x) \, \vev{  \sigma  \,  v }  dx $   which defines the denominator of  $\,k$, as has been already discussed. 
Then $\, k = 2 \, \xi(x_f) / ( \xi(x_f) + \xi(x_X)  )$ which as we have verified, following  our numerical procedure,  approximates fairly well  the numerical result for the same quantity. 
Notice that in this approximate form  the factor $k$ is a function of the ratio $ \xi(x_X) / \xi(x_f) $ which can be easily expressed, using Eq. (\ref{ratior3}), in terms of the parameters $b, c, \Delta_X $ 
and the freeze-out temperature $T_f$. 
The results displayed in Table \ref{tablecharter} show that the semi-analytic scheme  is successful in reproducing satisfactorily the numerical results with an accuracy in the range  $ \sim 15 \%  $. 
Therefore although not as accurate as in the conventional case, where  semi-analytic schemes yield better accuracies, the approximation used in the case at hand gives a fairly good picture of the relic density when dilaton effects are taken into account. 

Concerning  the factor $R$ a few remarks are in order. The values of $R$ obtained are of the order of $\, \sim {\cal{O}}{ ( 10^{-2} ) } $  and for fixed input values of $\, a, b$ decrease with increasing the mass $\,m_{{\chi} } $.  For high values of $\,m_{\tilde{\chi} } $, above a few  TeV, $\, R$ can become as small as 
$\, \sim {\cal{O} }{ ( 10^{-3} -10^{-4} ) } $. For small masses, smaller than $ 50 \, \GeV$ or so, the exponent in  Eq. (\ref{rfactor}) may not be small enough, being overwhelmed by the presence of 
$\, \xi^{-1}(x_f) $ in Eq. (\ref{rfactor}). This occurs because in these cases the point $\, x_X = T_X / m_{{\chi} } $, which sets the lower limit of the integration in the exponent of 
Eq. (\ref{rfactor}), becomes larger getting closer to the upper limit of the same integration. In these cases therefore  the factor $\, R$ may be larger than unity causing  no-reduction but enhancement instead to the relic density, by a factor which is $\, \sim {\cal{O} }{ ( 2 ) } $.  

For supersymmetric models, which we discuss in the following section, the reduction factor causes dilution by factor of order  ${\cal{O}} (5 \, - \,50) \,$, the smaller (larger) corresponding to lighter (heavier) neutralino masses.  The amount of the dilution depends on the particular SUSY inputs and therefore it is important that we scan the supersymmetric  parameter space to delineate regions compatible with the current accelerator and cosmological data, including those from direct dark matter searches. 
This we do in the following section. 
%%%%%%%%%%%%%%%%%%%%

\section{PREDICTIONS OF THE CMSSM}
As it was explained in the previous section the dilation terms may have significant impact on the predicted value for the 
neutralino relic density. In general, as shown in the previous section, for a specific point of the parameter space, the neutralino DM abundance is reduced. 
The effect of the reduction factor is plotted  in Fig.~\ref{fig1}, where in the context of the CMSSM we present examples of the $(m_0,M_{1/2})$ parameter space for $\tan\beta=5$ and $10$ respectively, taking
the trilinear coupling   $A_0=0$.  The purple thin lines, that they cross almost vertically the plane, mark the magnitude of this reduction factor  in the relic density.
Along the dotted purple line on the far left this factor is $0.1$, along the solid line it is $0.06$ and finally on the dashed line on the far right this is $0.006$. 

Turning to other details in these figures, the thin yellow shaded region on left is excluded 
by the LEP2 searches (chargino mass bound). The dark red region, of triangular shape, at the bottom is excluded because there the 
stau is the LSP. In the hatched grey regions at the upper left corner, whenever they appear, the EW symmetry breaking is not successful. 
The very thin blue strip that lies above the stau LSP region having also a segment that follows closely the edge of the EWSB forbidden region,  
delineates  regions that are compatible with the WMAP7 \cite{7yrwmap} data {\it if  dilaton effects are ignored}. On the other hand,
the dark green region is the WMAP7 compatible region in the presence of the  dilaton, while within the light green region 
the neutralino relic density is smaller than this bound. 
One immediately notices the significance of the new dilaton terms 
in the Boltzmann equation, by the resulting enlargement of the cosmologically favored regions. 
In the figures displayed the blue dashed-dotted curve denotes the light Higgs bound from LEP2, $m_h=114$ GeV.
The part of the parameter space on the left of the solid black line, marked by LHC, is already excluded by LHC experiments~\cite{lhc_last}. However large portions of the cosmologically allowed regions have 
been now moved to regions (dark green) of the parameter space that are not excluded by the current LHC data. 

Recent CERN-LHC data from Higgs searches provide additional constraints. The absence of pseudoscalar decay $ A \rightarrow \tau^+ \tau^- $ in ATLAS and CMS experiments puts  bounds on the pseudoscalar mass $\, m_A$ for given  $\, \tan \beta  $ ~\cite{CMSATLAS}. In particular,
 for large  $ \tan \beta \simeq 50 $ values  $ m_A \leq 450 \GeV$ are excluded. However this bound is not relevant in constrained supersymmetric scenarios, like CMSSM and mSUGRA, due to the fact that the masses of the Higgs bosons, but the light one, are quite heavy unless the SUSY breaking parameters are confined in the region of very small $\, m_0, \, M_{1/2}$ values. Therefore, this bound is practically inoperative in our analysis. Recall that 
large  $ \tan \beta \simeq 50  $ values allow rapid neutralino dark matter annihilations via the pseudoscalar A-resonance in the conventional mechanism where dilaton couplings to DM are neglected. This region has been phenomenologically studied recently, in view of the on-going LHC experiments, since the channel $ pp \rightarrow b+A \,(\mathrm{or}\, H) \rightarrow b \,  \mu^+ \mu^- + X$ allows for amplification of the signal to background-ratio  providing unique possibilities for measuring the Higgs boson masses and decay widths \cite{Baer:2011ua} at CERN LHC experiments. 

%%%%%%%%%%%
\begin{figure}[t!]
\begin{center}
\includegraphics[width=0.475\textwidth]{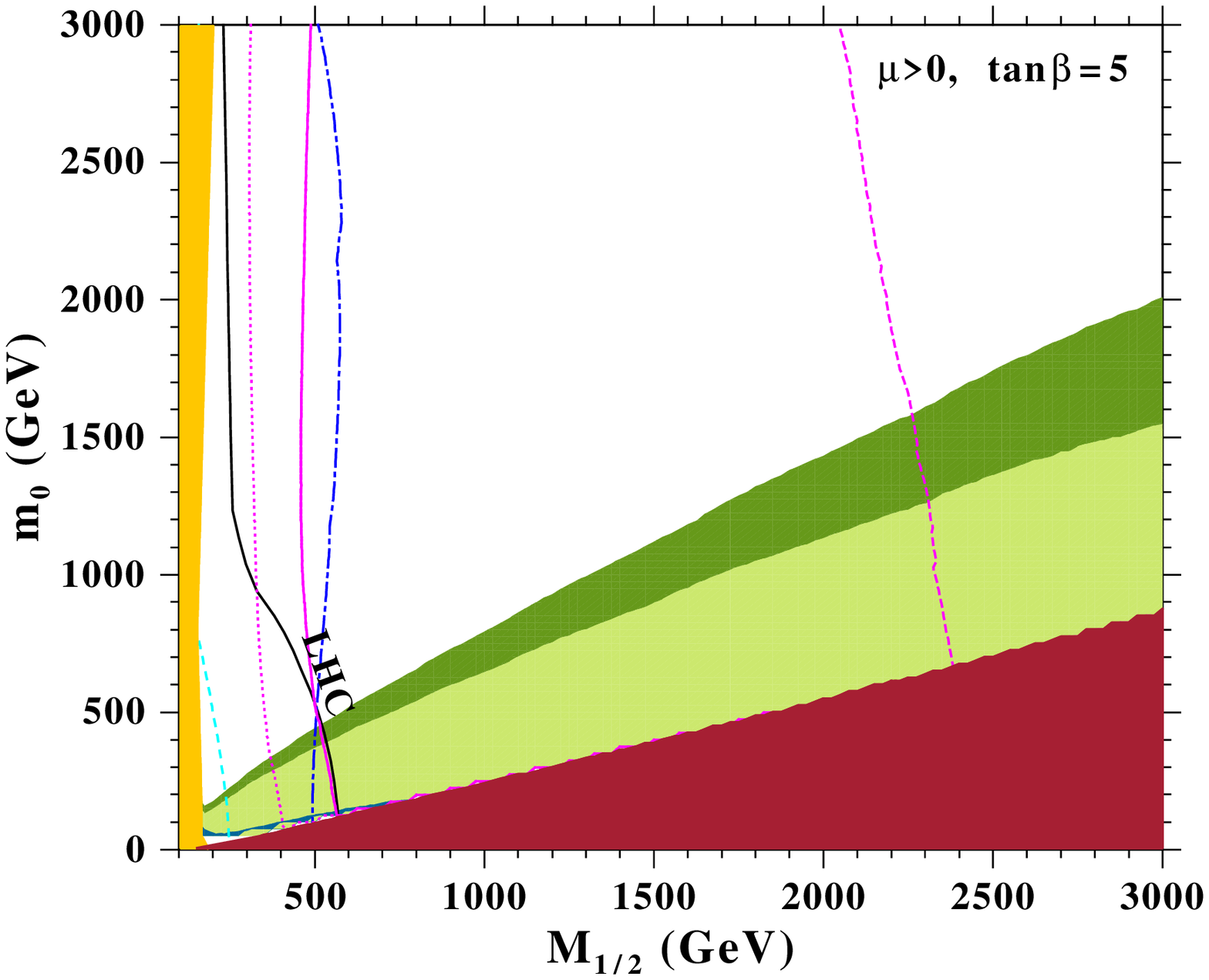}
\includegraphics[width=0.475\textwidth]{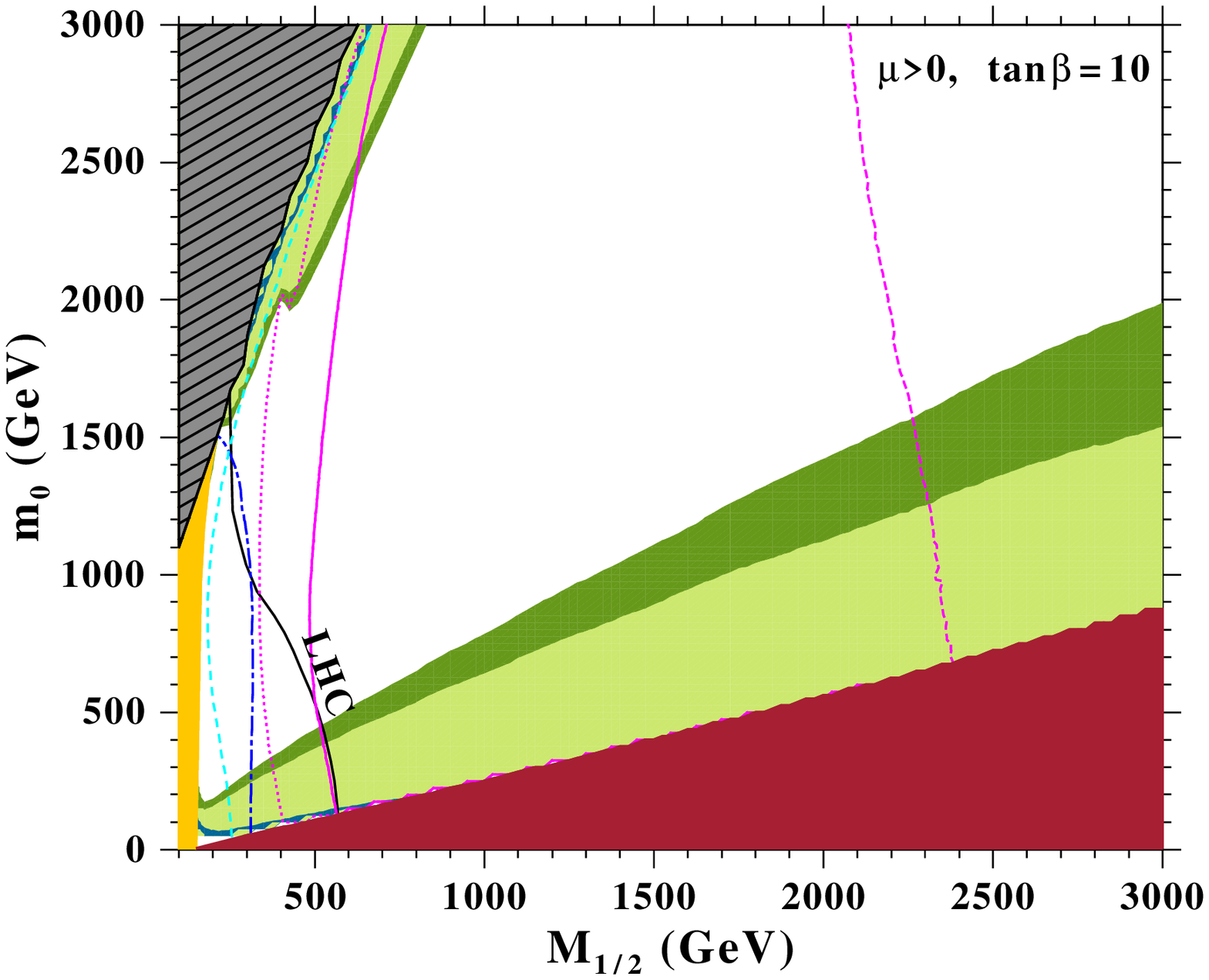}
\end{center}
\caption[]{
The $(m_0,M_{1/2})$ parameter space for $\tan\beta=5$ and 10, assuming $A_0=0$. The lines and regions are as described in the main text. }
\label{fig1}  
\end{figure}
%%%%%%%%%%%%%%%%%%%%%%%%%%
%%%%%%%%%%%
\begin{figure}[t!]
\begin{center}
\includegraphics[width=0.475\textwidth]{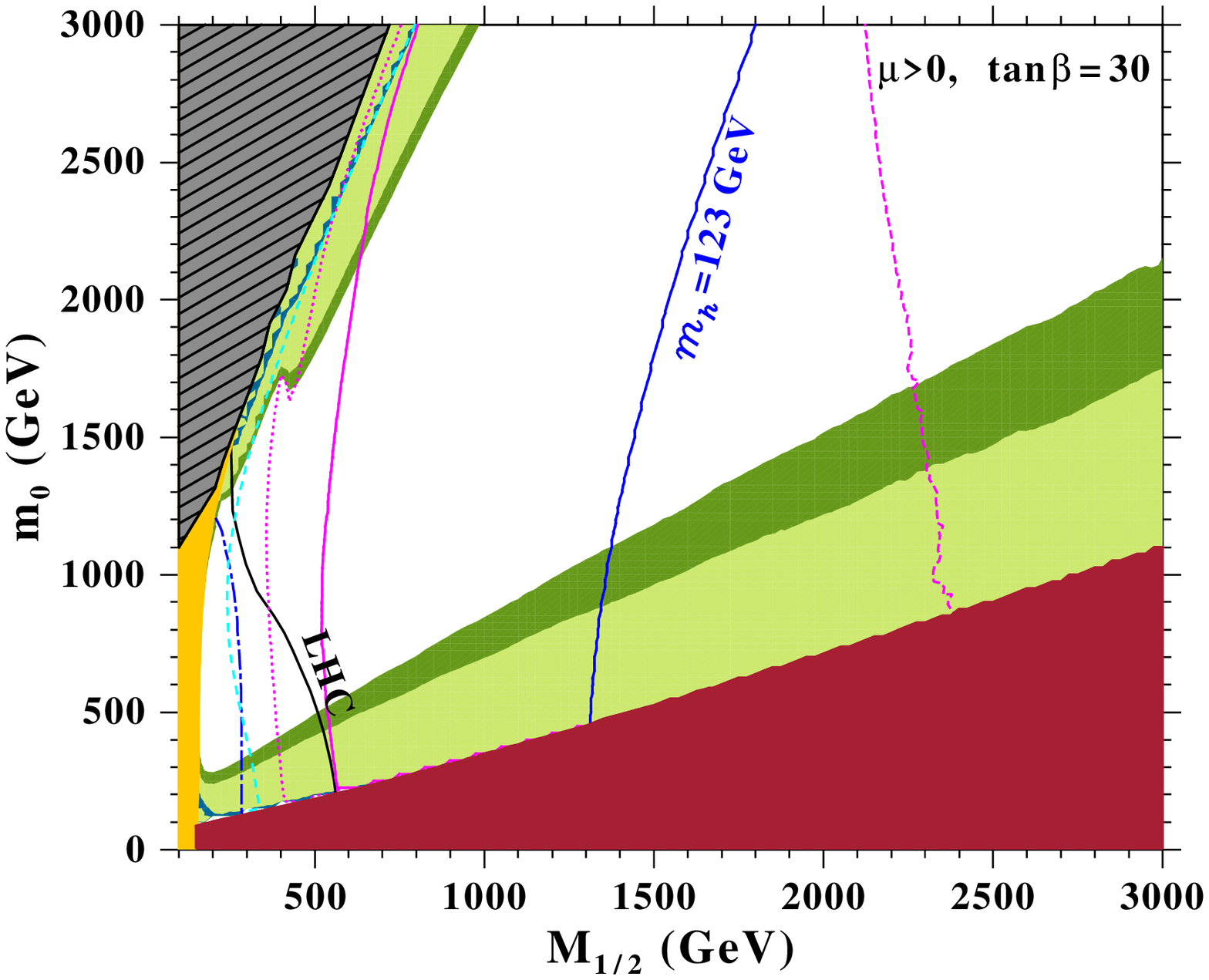}
\includegraphics[width=0.475\textwidth]{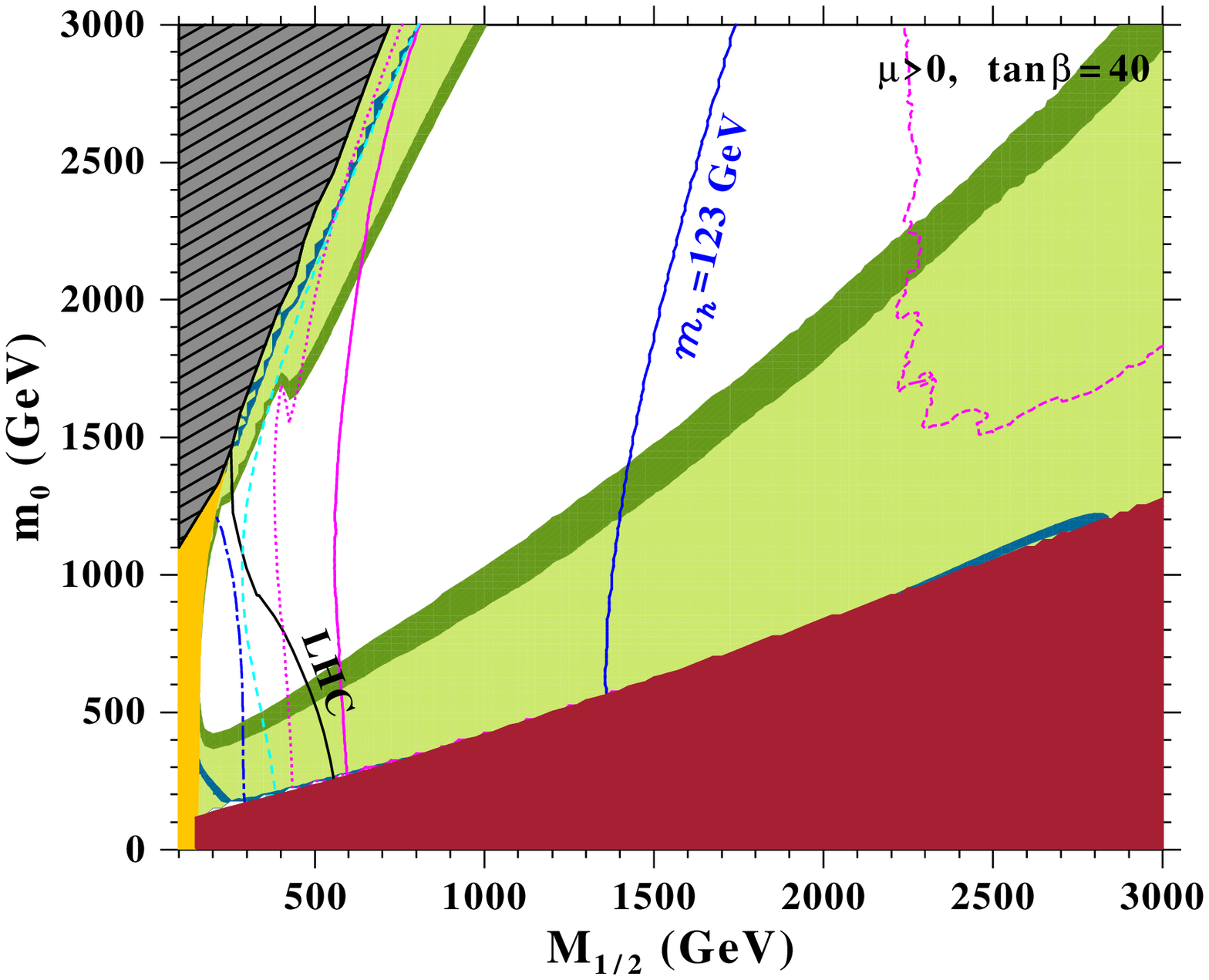}
\end{center}
\caption[]{
Like the Fig.~\ref{fig1} for $\tan\beta=30$ and 40.
}
\label{fig2}  
\end{figure}
%%%%%%%%%%%%%%%%%%%%%%%%%%

As for the light Higgs-boson, both  LHC experiments have reported possible evidence of a Standard Model like Higgs  
 in the  mass  range 123 -- 128 GeV~\cite{higgs_dec}. Thus in the figures presented in this paper we plot the curve  along which $m_h=123$ GeV, bearing in mind that in the regions of the CMSSM that we are studying the light CP-even Higgs boson accurately mimics the Standard Model Higgs. 
In  Fig.~\ref{fig1}, for $\tan\beta=5$ and $10$ respectively, all the displayed parameter space   yields 
Higgs masses smaller than 123 GeV. For larger values of $\tan\beta$, displayed in the other figures, we notice that large part of the parameter space 
is compatible with $m_h>123 $ GeV, which lies in the region right to the blue curve that illustrates  the $m_h=123 \GeV $ Higgs boson mass. 
Actually, in Fig.~\ref{fig2}  presented  here, with $M_{1/2}<3$ TeV, the Higgs boson is always lighter than 124--125 GeV. 
Note that a light Higgs boson with mass larger than 120 or 123 GeV  it is not easy to 
be compatible with cosmologically favored regions of CMSSM for all the values of $\tan\beta$, as was discussed 
recently~\cite{higgs_ph}. But, taking into account the dilaton effects on the computation of the neutralino relic density,
we notice that large parts of the parameter space become compatible both to WMAP bound and to Higgs mass larger that 123 GeV,
especially for $\tanb > 25$.
In particular, for $\tan\beta$ in the range 25 to 40 large portions  of the ``dilaton deformed" coannihilation corridor have Higgs
masses larger than 123 GeV, as can be seen in Fig.~\ref{fig2}. 
For $\tan\beta$ in the range 40 to 48 there are such regions  that are the ``dilaton deformation" of the pseudo scalar  Higgs boson  funnel,
which is the blue shaded region as shown in Fig.~\ref{fig3} left. It is obvious, that larger regions are compatible to these Higgs boson masses if one assumes that neutralino 
relic density is smaller than the WMAP bound, which allows for contribution to dark matter density from other particle species.

%%%%%%%%%%%
\begin{figure}[t!]
\begin{center}
\includegraphics[width=0.475\textwidth]{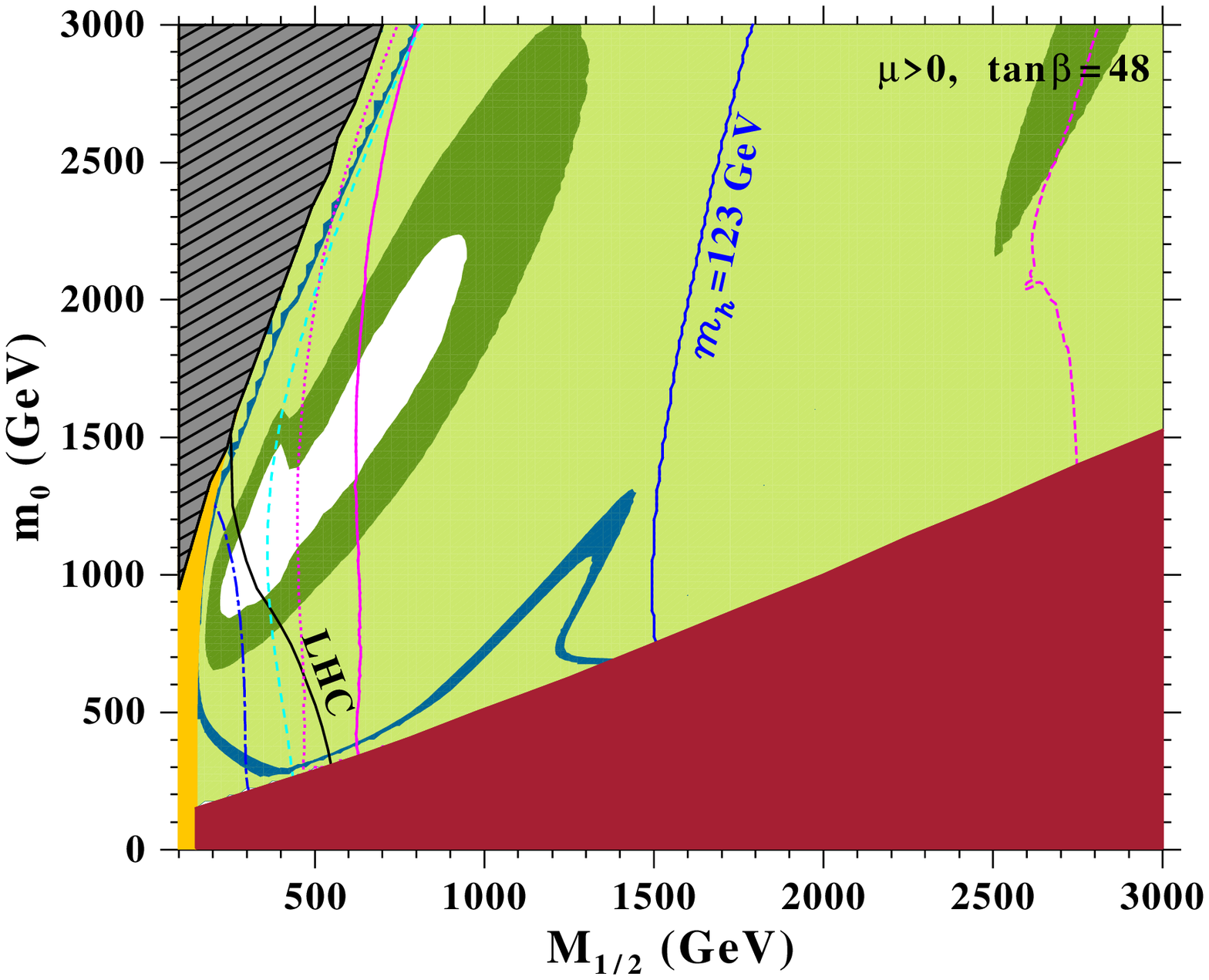}
\includegraphics[width=0.475\textwidth]{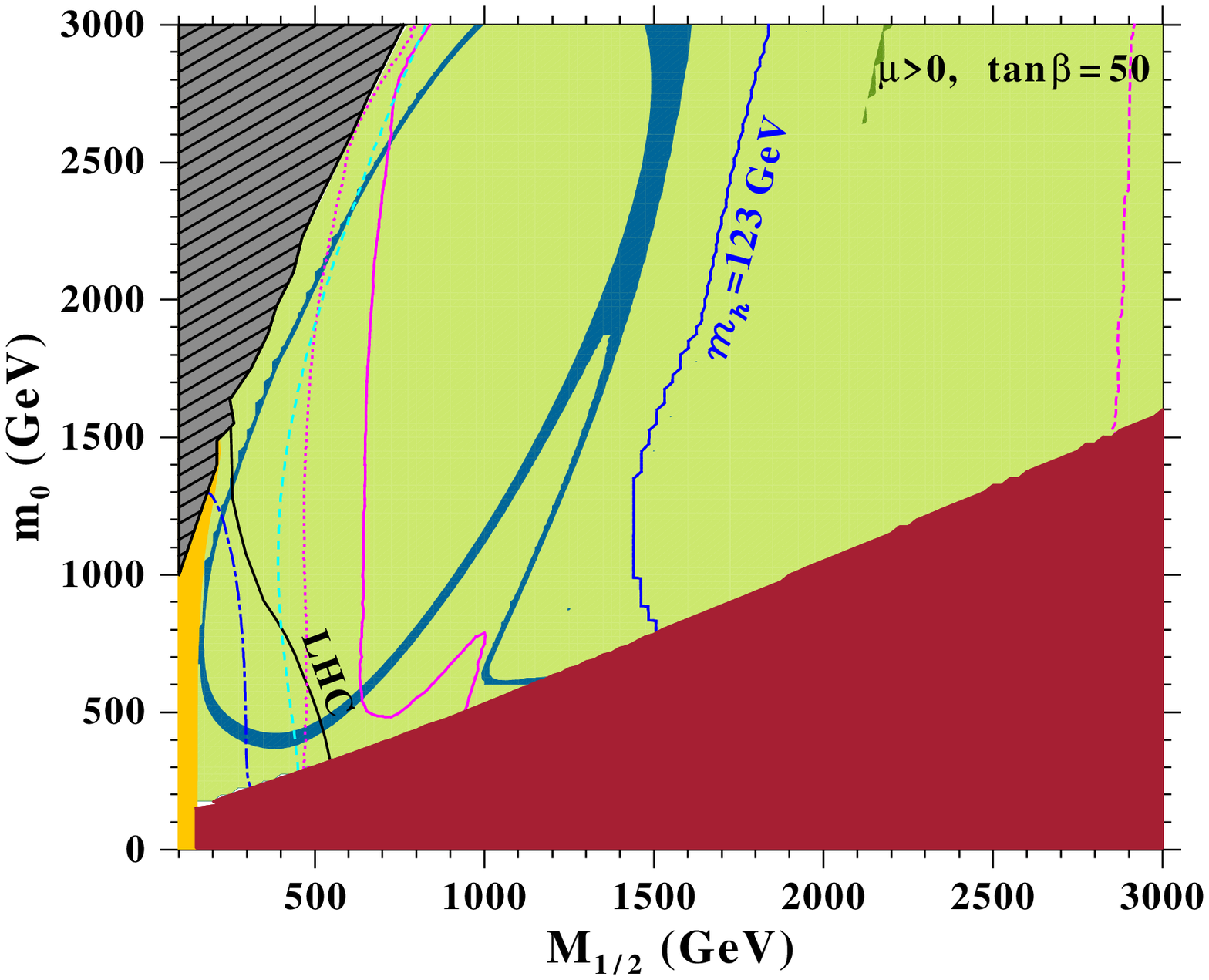}
\end{center}
\caption[]{
Like the Fig.~\ref{fig1} for $\tan\beta=48$ and 50.
}
\label{fig3}  
\end{figure}
%%%%%%%%%%%%%%%%%%%%%%%%%%

%%%%%%%%%%%%%%%

The cyan dashed line, on the left of each figure, delineates  the bound set by XENON100 direct detection  measurement~\cite{xenon100}. The excluded region is on the left of this 
curve and includes the focus point region in cases where the dilaton effects are ignored. 
It's worth noticing that in order to apply the neutralino DM direct detection limit, 
one must take into account the hadronic uncertainties that dominate the neutralino-nucleon interaction. 
The dominant uncertainty stems from the strange quark density in the nucleon $\bra{ N}  \bar{s}s   \ket{N}$,
which reflects the uncertainty in the  pion nucleon sigma term $\Sigma_{\pi N}=1/2(m_u+m_d) \bra{ N}  \bar{u}u +  \bar{d}d  \ket{N}$~\cite{hadron}.
Various estimates of the $\Sigma_{\pi N}$ vary  from $64\pm8$ MeV to the more conservative $50\pm14$ MeV~\cite{sigvalue}.
In this analysis we have adopted the value $\Sigma_{\pi N}=50$ MeV.
%%%%%%%%%%%%%%%%%%%%%
Even with this conservative value a significant portion of the focus point corridor, especially for 
large $\tan\beta $, is excluded. The reason is that large part of the WMAP7 cosmologically allowed region, in the absence of the dilaton, is within the XENON100 exclusion area, as is seen, for instance in Fig.~\ref{fig3}. 
This happens for values of $M_{1/2}$ up to 1000 GeV, that 
correspond to $m_\chi \lesssim 450 $ GeV, where the XENON bounds are fully applied.\footnote{The published XENON100 bound applies for $m_\chi$ up to 1 TeV, although an extrapolation of the bound can be performed.}. 
Interestingly enough, the dilaton effects reshape the WMAP allowed region in such a way that it now lies in a region where 
XENON100 bounds are evaded. Hence the focus point region is a viable possibility in this scenario. 

Concluding this section, we showed that the dilaton coupling to DM may dramatically modify the predictions of the supersymmetric models. 
In the regions of the CMSSM parameter space that is accessible to the LHC experiments up to the $\sqrt{s}=14 \TeV$, the combination of WMAP7, the Higgs boson searches at LHC and 
the direct detection experiments like XENON100, put new  phenomenological restrictions \cite{higgs_ph}. 
However in the presence of dilaton driven terms  in Boltzmann equation these regions are deformed in such a way that much larger parts of the
supersymmetric parameter space are compatible to cosmological bounds, to XENON100 and to Higgs boson masses in the region 123--128 GeV,
as it is indicated from the recent LHC data. This opens new directions for further phenomenological analyses.

\section{INDIRECT DARK MATTER SEARCHES:  $\gamma$-RAY FLUXES}
Indirect DM searches, dedicated  to detect  potential   $\gamma$, neutrino or positron 
fluxes,  produced by the DM particle annihilations,   have been intensified during last years. 
Especially, photon data   collected from various  regions of the Universe, like the centers of galactic halos where a large density of DM is clumped,
if they are  above the known backgrounds, may  signal the presence of DM.
Currently these can be detected by the new generation of detectors whose experimental precision has been 
 increased significantly, like Fermi-LAT or HESS, and ground-based experiments expected to operate in the future, like the Cherenkov Telescope Array (CTA). 

In this part using the $\gamma$-ray data, delivered by Fermi-LAT \cite{fermi,fermi2,fermi3,dsphfermi} and HESS 
\cite{HESS} experiments, and taking into account the projected sensitivity of the CTA detector \cite{CTA}, which is scheduled to operate in the future, we  briefly discuss the consequences  of the dilaton scenario for the aforementioned indirect DM searches, and investigate how  dilaton  modifies the conventional picture. 
In the  analysis of the previous section  we have mainly concentrated on predictions of supersymmetric models, namely the CMSSM, and saw that the dilaton may dilute the relic density altering the constraints imposed by the WMAP data, 
allowing smaller neutralino annihilation rates. 
This in turn, implies that supersymmetric models are not expected to be tightly constrained  by indirect Dark Matter search 
data. However given the fact that in the near future new more precise experiments  will start delivering  data,  
improving the limits put on the DM annihilation cross section, it is important that we consider current and proposed  limits put by these experiments too. 
It should be stressed  that such an analysis is important not only for supersymmetry, but for any model that predicts the existence of Cold Dark Matter that annihilates to ordinary matter, irrespectively of the nature of the DM particle and the particular mechanism through which annihilation into standard model particles proceeds. 

Upper bounds on the WIMP pair annihilation cross section can be obtained from the $\gamma$-ray flux, due to 
the  WIMP  annihilations in various regions of the galactic DM halo
\cite{Bergstrom:2010gh,Abazajian,Donato:2011pe,Cholis:2012am}. 
This flux is proportional to the non-relativistic limit of  
the thermally averaged annihilation cross section times the relative velocity $ \vev{ \sigma  \,  v }  $, the energy spectra $ \, d N / d E_\gamma $  originating from DM annihilation into final states 
$ q \bar{q}, \tau^+ \tau^{-}, \mu^+ \mu^{-}, W^+ W^{-}, \, Z Z,\, Z h $ etc, and the $J$-factor that depends on the DM profile. 
In particular, the gamma-ray signals from dwarf spheroidal galaxies observed by Fermi-LAT \cite{dsphfermi} exclude 
WIMPs with masses $\, m_{\chi} \lesssim 40 \, GeV$ annihilating into $\, b \bar{b}, \tau^+ \tau^{-} $, 
while HESS observations provide constraints for heavier WIMPs~\cite{Abazajian,hessgc}.
%%%%%
In order to  study  the indirect DM searches, in conjunction with the dilaton scenario, we
will concentrate on various  $\gamma$-ray data originating either from the galactic center (GC) or the dwarfs spheroidal  (dSph) galaxies.
In Fig.~\ref{fig_sigm} we present constraints  from Fermi-LAT, HESS and CTA (projected) on the the $ ( \vev{\sigma \, v}, m_\chi ) $ plane,
assuming that the dominant DM pair annihilation goes through the $b\, \bar{b}$ channel and all other channels have  negligible contribution. 
We also assume that since the WIMPS annihilating in the galactic halo today are very non-relativistic, the annihilations are essentially pure S-wave.
Both of these assumptions are realized in the major portions of the parametric space of 
supersymmetric models but evidently can hold in other cases too. 
The Fermi-LAT experiment can detect $\gamma$-rays in the energy range 20 MeV to 300 GeV.  For this analysis we use the  
Fermi-LAT data recorded between   August 4th, 2008 and April 20th, 2012. We perform a data analysis using the package  ScienceTools~\cite{tools} as it is suggested by the collaboration. As it was done in ~\cite{dsphfermi} we are using data from ten dSph galaxies: Bootes I, Carina, Coma Berenices, Draco, Fornax, Sculptor, Segue 1, Sextans, Ursa Major II and Ursa Minor. Reproducing the statistical analysis in  ~\cite{dsphfermi}, and using the J-factors for dSph galaxies used in this analysis, we calculate the combined  95\% CL upper limits shown in Fig.~\ref{fig_sigm} for the joint likelihood of these ten dSph. These are in agreement with Fig. 2 presented in   ~\cite{dsphfermi}.  
Looking into the physics of this constraint, one can see that practically assuming that the dominant annihilating channel is the $ b\, \bar{b}$ only models with low $m_\chi$ are constrained. In fact, in the presence of the dilaton, models with $m_\chi$ up to 25 GeV are excluded. This conclusion is 
drawn using the WMAP compatible bands that are plotted in  Fig.~\ref{fig_sigm}. Within the light blue horizontal band 
of the $ ( \vev{\sigma \, v}, m_\chi ) $ plane the relic density is within the WMAP limits  in the conventional case, i.e. ignoring  dilaton effects, and  the green strip bending downwards as mass increases designates the WMAP compatible region  when the dilaton is present, as was described in  section III. 
Notice that for $\, m_{\chi} \lesssim 30 \, \GeV $ the dilaton 
predictions requires larger cross sections than in the conventional scenario and the green strip lies above the corresponding blue one. 
In fact, in this low $\, m_{\chi}$ regime the factor $\,R$ causes enhancement, as has been explained in the previous sections.

%%%%%%%%%%%
\begin{figure}[t!]
\begin{center}
\includegraphics[width=12cm]{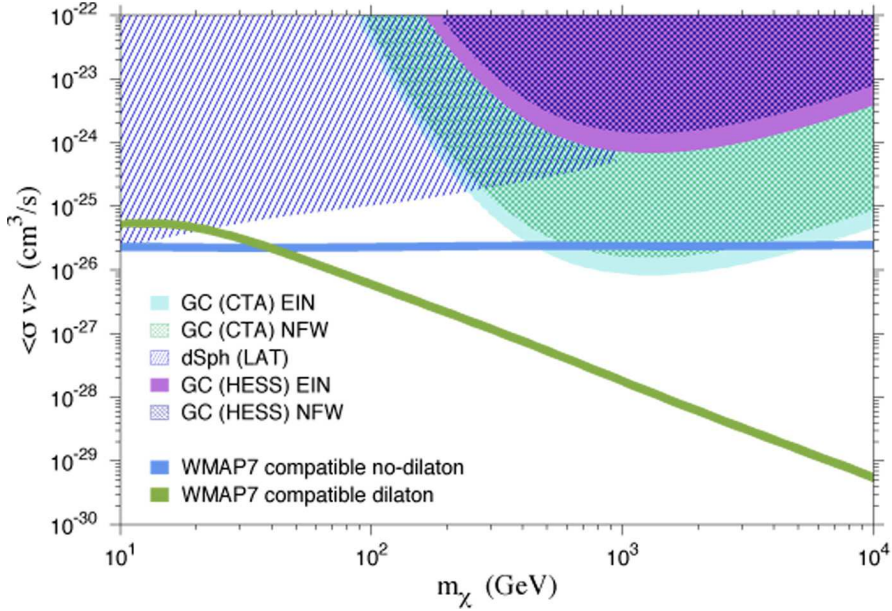}
\end{center}
\caption[]{
Bounds of various indirect gamma ray experiments  on the $ ( \vev{\sigma \, v}, m_\chi ) $ plane. The profires  EIN = Einasto  and 
NFW = Navarro-Frenk-White are used.  
}
\label{fig_sigm}  
\end{figure}
%%%%%%%%%%%%%%%%%%%%%%%%%%

In the same figure  we have also plotted the regions that are excluded by HESS measurements, based on $\gamma$-ray fluxes originated 
from the GC~\cite{hessgc}, assuming an observation window of 10 degrees around  this. 
The HESS instrument is sensitive to photons in the energy range  300 GeV to 30 TeV. In order, to 
delineate the  exclusion region at 95\% CL on the $ ( \vev{\sigma \, v}, m_\chi ) $  plane, we divide the  HESS energy range into 30 bins evenly spaced  on a logarithmic scale. For the calculation of the supersymmetric signal we include the monochromatic, 
the continuum and inverse Compton scattering components as they are described in detail  in~\cite{eos_gc}. We use two profiles 
for the halo   density around the GC: the Navarro-Frenk-White (NFW)~\cite{NFW} and Einasto~\cite{einasto},
with the choice $\rho=0.3  \GeV / \mathrm{cm^3}$, for the local DM density,  and $ R_0=8.5 $ Kpc for the solar 
distance to the GC  (more details on  the halo profile parameters and the methodology that we follow  can be found  in~\cite{eos_gc}). 
The background contribution consists of the diffuse galactic emission, the isotropic extragalactic contribution and resolved point source contribution.
 We model these background contribution using the latest models that are available from the Fermi-LAT collaboration~\cite{bckg}. 
 We construct a  $\chi^2$ defined by $\chi^2=\sum_i S_i^2/(S_i+B_i)$, where $S_i$ and $B_i$ are the signal and the background 
counts per bin, respectively.  Then we compute the corresponding 95\% CL exclusion limit based on this $\chi^2$ which gives
 a curve  quite similar to the bounds presented in Fig. 4 of  ~\cite{hessgc}.
In the same manner we can draw the region on the $ ( \vev{\sigma \, v}, m_\chi ) $  space  that can be reached by the future 
CTA detector~\cite{CTA}. We use a similar $\chi^2$ analysis, but calculated with an observation time and effective area for this 
detector as  given in Table II of~\cite{Bergstrom:2010gh}, following the  discussion presented in section V of this paper.
The projected excluded region by CTA is again presented in Fig. \ref{fig_sigm} for the NFW and the Einasto profiles. 
 This is similar  to  Fig. 2 of ~\cite{Bergstrom:2010gh} when one uses the same units and divides the cross section by  $ m_{\chi}^2 $. 
Based on these, we 
are expecting that CTA detector can probe annihilations with values of  $  \vev{\sigma \, v}$ of the order of $10^{-26} \mathrm{cm^3/s}$ 
that overlap, in the conventional model, with the cosmologically favoured WMAP region for $\, m_{\chi}$ masses in the range 
$\, m_{\chi} \simeq 400 \, \GeV - 5 \, \TeV $ as can be seen from Fig.~\ref{fig_sigm}. 
On the other hand, in the context of the dilaton cosmologies, discussed in previous sections, much smaller values of 
$  \vev{\sigma \, v}$ are compatible to the WMAP data and therefore in the presence of the dilaton  studied in this work, it seems that even with the CTA scheduled performance it will be hard to derive a useful constraint.

%%%%%%%%%%%
\begin{figure}[t!]
\begin{center}
\includegraphics[width=0.475\textwidth]{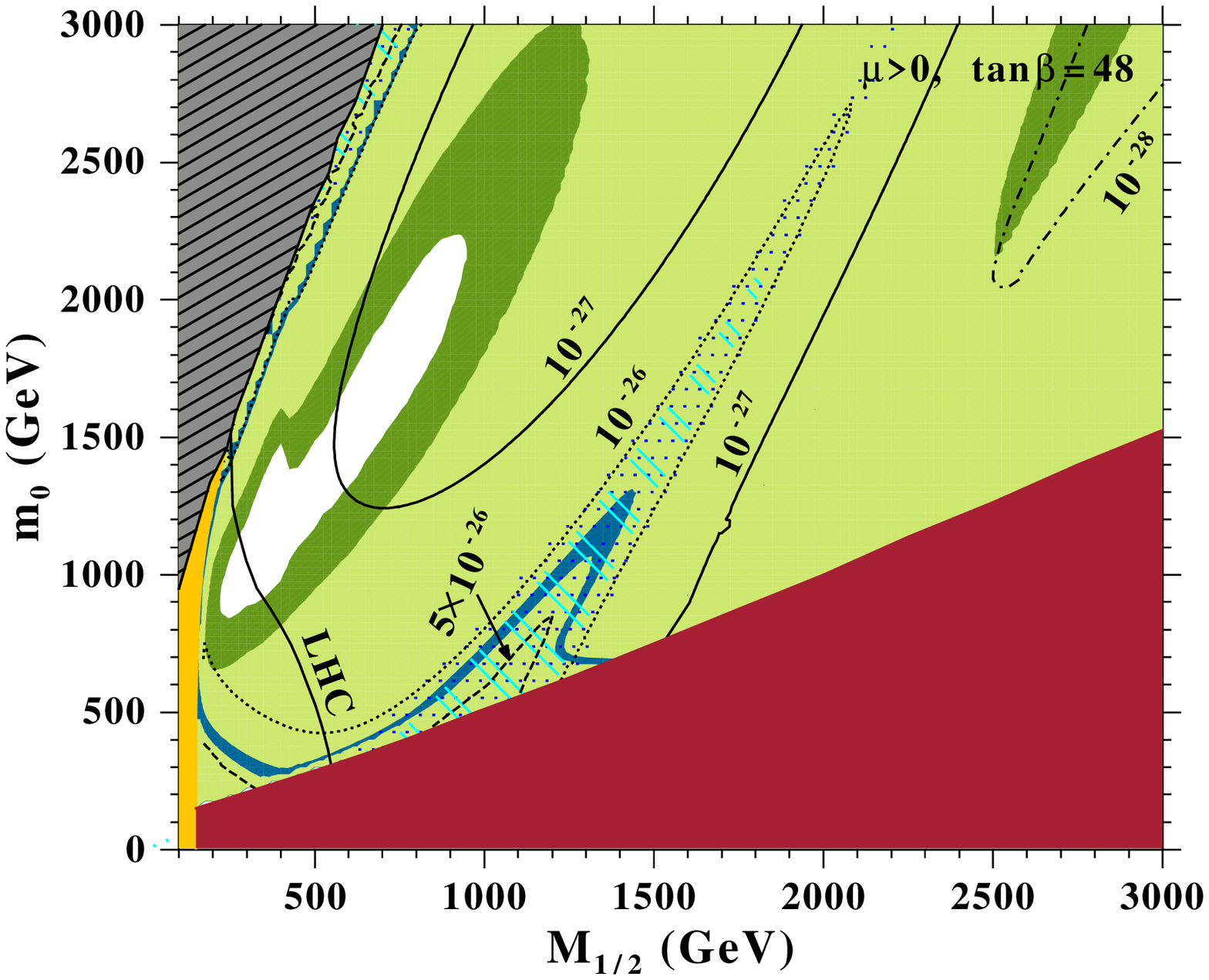}
\includegraphics[width=0.475\textwidth]{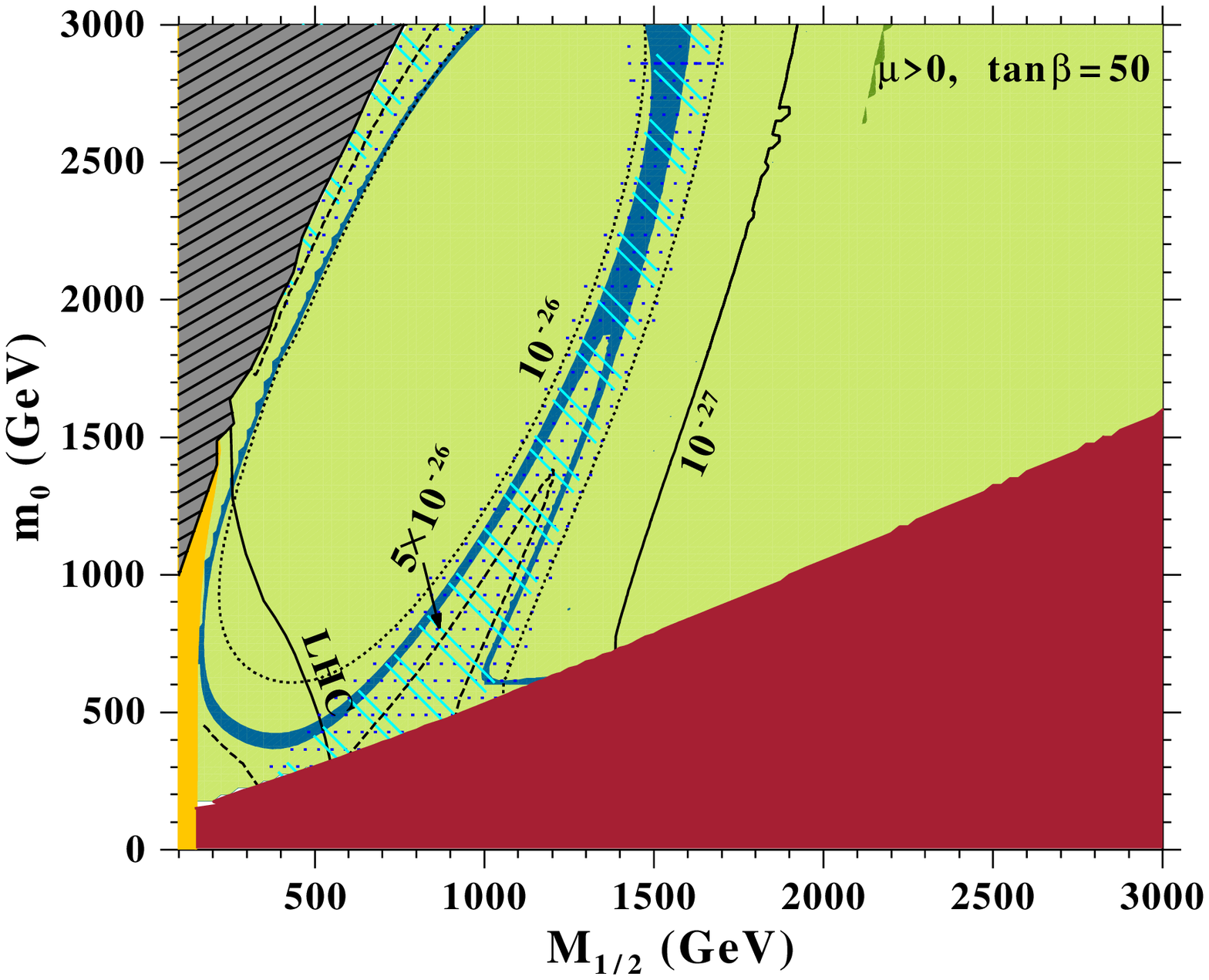}
\end{center}
\caption[]{
Like Fig.~\ref{fig3}, including contours of constant $\vev{\sigma \, v}$, in  $\mathrm{cm^3/s}$.
Also  shown are the regions expected to be excluded by CTA experiment from the projected data collected from measurements around the GC, as  described in the text.
}
\label{fig4}  
\end{figure}
%%%%%%%%%%%%%%%%%%%%%%%%%%

Passing  to  the case of CMSSM, in 
Fig.~\ref{fig4} we present in the  $(m_0,M_{1/2})$ parameter space, for $\tan\beta=48$ (left plane)  and 50 (right plane),  
contours of the total pair annihilation cross section $  \vev{\sigma \, v}$  
 for values of it $10^{-28},\, 10^{-27},\, 10^{-26}$ and $5\times 10^{-26}\, \mathrm{cm^3/s}$ respectively. 
We notice, as  expected from the previous discussion on Fig.~\ref{fig_sigm}, that the parts of the parameter space that are cosmologically favoured in the conventional model are compatible 
with values  $  \vev{\sigma \, v} \sim 2.5 \times 10^{-26} \mathrm{cm^3/s}$. 
Therefore these regions can be probed by the CTA expected measurements. However, these measurements will be unable  to probe models with relic density calculations affected by the dilaton effects. 
To illustrate this in greater detail we  also mark on these figures the regions of the parameter space likely to be constrained by the CTA from the GC (10 degrees window) for the NFW (cyan inclined lines) and the Einasto (blue horizontal dots) profiles. 
These regions surround the cosmologically allowed regions of the conventional model, 
 but stay apart from  the corresponding WMAP allowed regions in the presence of the dilaton,  being in qualitative agreement with the findings regarding Fig.~\ref{fig_sigm}. Note also that in this case 
for the calculation of this projected bound we have used all the neutralino pair annihilation channels occurring  in the CMSSM model and not just the $b\, \bar{b}$ as we did before when considering the case depicted in Fig.~\ref{fig_sigm}. 
The annihilation channel  $\, b \bar{b}$ still dominates  in the major portion of the parameter space, 
except the focus-point region, where $\, W^+ W^{-}$ dominates, and the neutralino-stau coannihilation region where $ \tau^+ \tau^{-} $ is the dominant channel. 

The conclusion reached in this part is that the cosmologically favored regions of the parameter space of the CMSSM can probably beyond reach of  future planned indirect $\gamma$-ray DM detection experiments, due to the dilaton effects which dilute  the DM relic density preferring smaller annihilation cross sections. Although we have demonstrated this for high values of $\tan \beta$ this holds true for smaller values  of the parameter $\tan\beta$ as well. In particular in these cases, the regions compatible with the cosmological bounds for the relic density have cross sections  
$\, \vev{\sigma  \,  v } \sim 10^{-28} \, \mathrm{cm^3 / s} $, or less, well below the sensitivities of current and planned future experiments.

\section{Conclusions}
In this paper we have shown that the dilaton dynamics during early eras, long before Nucleosynthesis, in conjunction with its coupling to Dark Matter may have dramatic consequences for the predicted Dark Matter relic density in the popular supersymmetric models. 
Modeling the dilaton evolution to be that dictated by exponential type  potential, occurring in quintessence scenarios, inflation models  and string theory,  the ordinary predicted DM density may be diluted by large factors ranging from  $\, {\cal{O}} (5) \,$ to $\, {\cal{O}} (50) \,$. This dilution mechanism is consistent with the absence of dilaton couplings to ordinary matter ( hadrons ), in the continuity equations, but it affects DM relics since dilaton dominates over radiation during and after  DM decoupling. 
This allows for LSP annihilation cross sections, in the popular supersymmetric schemes, that are smaller by an order of magnitude or more. 
This however may imply smaller inelastic cross sections of the neutralino LSP with nucleons putting farther the potential of discovering supersymmetric DM at proposed direct detection experiments 
\cite{direct,xenon100}. 
We studied the predictions of the CMSSM in the light of LHC recent data and cosmological constraints arising from WMAP7 and XENON100 which puts the most stringent constraints of all  direct DM search experiments. We found that the allowed points cover broad regions that occupy different regions of the parametric space as compared to ordinary models. This opens new directions for further phenomenological studies and prolongs the viability of supersymmetric models. 

As for  indirect detection experiments, indirect searches of Dark Matter through antimatter production has stirred much interest the last three years. These data may be conditionally explained 
as DM annihilation in the galactic halo, generating the produced antiparticle flux. However more conventional interpretations exist that can explain the origin of these fluxes. The $\gamma$-rays measured by various experiments is another powerful tool for indirect detection of DM. In the cases studied in this work the smaller annihilation cross sections,  which are required in this dilaton scenario in order to satisfy the WMAP data, weakens the potential of discovering  neutralino DM in the galactig halo, through $\gamma$-ray current and future planned experiments, unless the detector performances are greatly improved.

\vspace*{.5cm}

\section*{Acknowledgments}
The work of A.B.L. is supported  by the European Union under the Marie Curie Initial Training Network ``UNILHC" PITN-GA-2009-237920 and RTN European Programme  MRTN-CT-2006-035505 HEPTOOLS and also co-funded by the European Union and Greek national funds through THALIS program.A.B.L. acknowledges also support by the UoA Special Research account. 
The work of V.C.S. is supported by Marie Curie International Reintegration grant SUSYDM-PHEN, MIRG-CT-2007-203189.

\end{document}